\pdfoutput=1
\documentclass[11pt,a4paper]{article}
\usepackage{geometry}
\usepackage{amsmath}
\usepackage{graphicx,subfigure}
\newcommand{\discont}[1]{[\![#1]\!]}
\DeclareMathOperator{\tr}{tr}

\newcommand{\raisedchi}{\chi}

\begin{document}



    \title{
    Capillary buckling of a thin film adhering to a sphere
    }

    \author{J.~Hure\textit{$^a$} \and B. Audoly\textit{$^b$}}    
    
\maketitle 

    \begin{abstract}
	We present a combined theoretical and experimental study of
	the buckling of a thin film wrapped around a sphere under the
	action of capillary forces.  A rigid sphere is coated with a
	wetting liquid, and then wrapped by a thin film into an
	initially cylindrical shape.  The equilibrium of this
	cylindrical shape is governed by the antagonistic effects of
	elasticity and capillarity: elasticity tends to keep the film
	developable while capillarity tends to curve it in both
	directions so as to maximize the area of contact with the
	sphere.  In the experiments, the contact area between the film
	and the sphere has cylindrical symmetry when the sphere radius
	is small, but destabilises to a non-symmetric, wrinkled
	configuration when the radius is larger than a critical value.
	We combine the Donnell equations for near-cylindrical shells
	to include a unilateral constraint with the impenetrable
	sphere, and the capillary forces acting along a moving edge.  A
	non-linear solution describing the axisymmetric configuration
	of the film is derived.  A linear stability analysis is then
	presented, which successfully captures the wrinkling
	instability, the symmetry of the unstable mode, the
	instability threshold and the critical wavelength.  The motion
	of the free boundary at the edge of the region of contact,
	which has an effect on the instability, is treated without any
	approximation.
     \end{abstract}



\section{Introduction}

\footnotetext{\textit{$^{a}$~Univ Paris Diderot, Sorbonne Paris Cit\'e, PMMH, UMR 7636 CNRS,
ESPCI-ParisTech, UPMC Univ Paris 06, F-75005 Paris, France}}
\footnotetext{\textit{$^{b}$~UPMC Univ Paris 06, CNRS, UMR 7190, Institut Jean Le Rond d'Alembert, F-75005 Paris, France}}

The buckling of thin plates has been studied for a long time, both
theoretically~\cite{platetimoshenko,timostab} and
experimentally~\cite{singer}.  Initially, the main motivation was to
avoid loads associated with catastrophic failure modes.  Recent
research efforts on thin plates and shells have been driven by
technological applications involving thinner and thinner
plates~\cite{geim}, by the idea that controlled buckling can provide
useful functionality~\cite{shim}, and by strongly non-linear phenomena
appearing far above the bifurcation threshold~\cite{bennypnas}. 
The wrinkling of
a semi-infinite elastic medium under finite compression, known as 
Biot's problem~\cite{Hohlfeld-Mahadevan-Unfolding-the-sulcus-2011,%
Cao-Hutchinson-From-wrinkles-to-creases-2011}, as well as the buckling
of a thin stiff film coated to a compliant
substrate~\cite{ChenHutchinson-Herringbone-Buckling-Patterns-of-Compressed-Thin-2004,%
AudolyBoudaoud-Thin-film-on-compliant-substrate-Part3-LargeCompression-2008,%
Cai-Breid-EtAl-Periodic-patterns-and-energy-2011} are just two
examples of classical problems in mechanical engineering whose
non-linear aspects have been well understood only recently.  We refer
the reader to~\cite{li} for a comprehensive review.

When the adhesion between a thin film under residual compression and a
thick substrate is relatively weak, buckling can take place along with
delamination, resulting in the formation of
blisters~\cite{gioia,vella}.  Buckling-driven delamination can lead to
various patterns which are affected both by the
mode-mixity~\cite{suo}, \emph{i.\ e.}\ the dependence of the
interfacial energy on the loading mode, and by the irreversibility of
the interfacial
fracture~\cite{AudolMode-dependent-toughness-and-the-delamination-of-compressed-thin00}.
In recent experiments, a simpler variant of the classical delamination
problem has been proposed, whereby the adhesion between the film and
the substrate is provided by capillary forces arising from a thin
liquid bridge~\cite{bico,hure2011}: capillary forces are reversible
and act like a self-healing interface crack.  These experiments can be
done
at the centimeter scale. 

\begin{figure}[tbp]
    \centerline{\includegraphics[width=.7\textwidth]{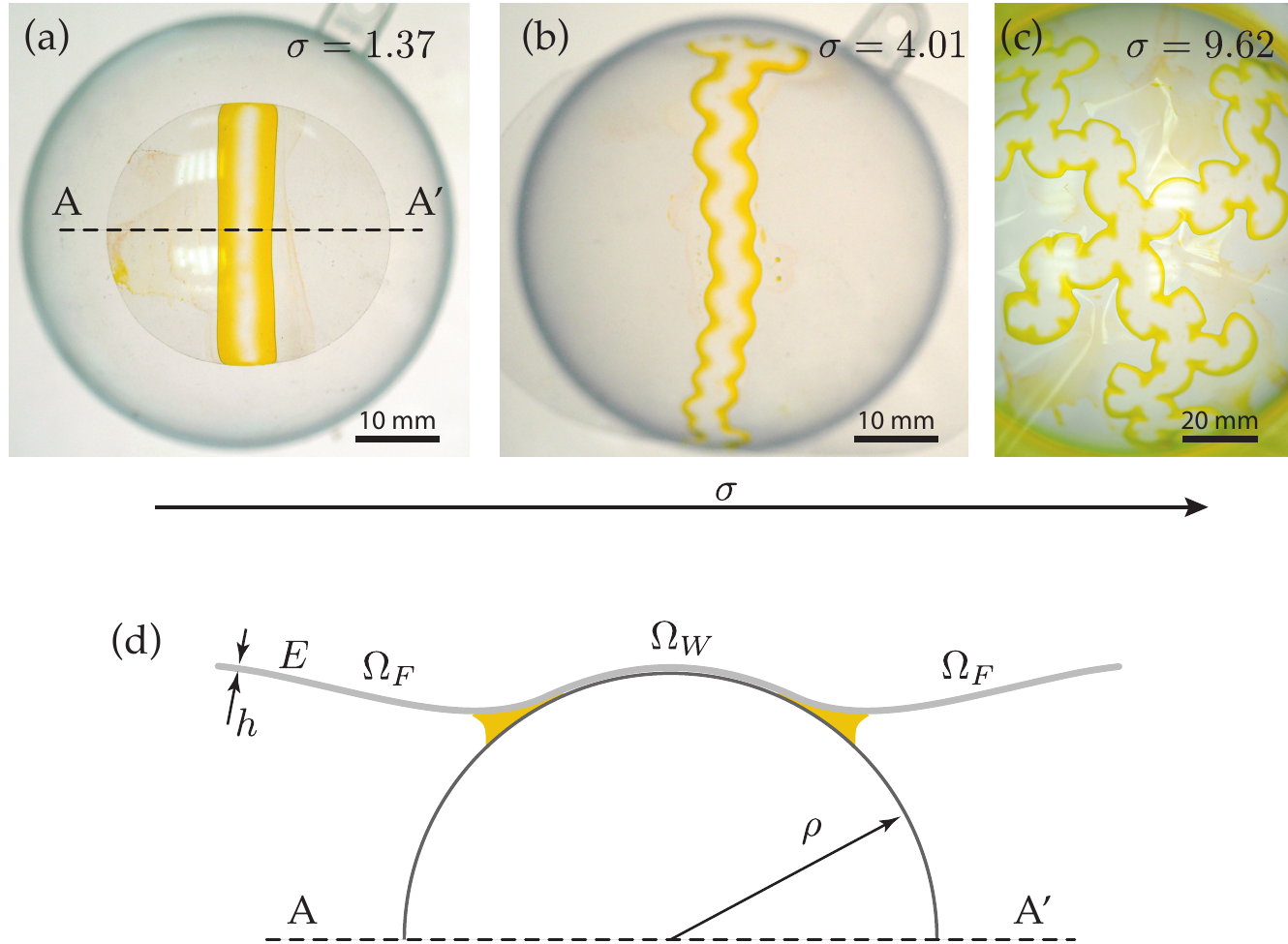}}
    \caption{Thin elastic films of thickness $h$, Young's modulus $E$
    and Poisson's ratio $\nu$ are placed onto rigid spherical caps of
    radius $\rho$ coated with a wetting liquid, with surface tension
    $\gamma = 22.4 \mathrm{mN.m^{-1}}$.  The liquid has been dyed in
    yellow to help visualization: the yellow region are the liquid
    meniscus.  Note that the interior of the region of contact appear
    uncolored as there is almost no fluid there.  Top view of three
    different experiments: (a) $E=2.8\,\mathrm{GPa}$, $h=30\,\mu$m,
    $\rho=25\,$mm (b) $E=2.6\,\mathrm{GPa}$, $h=15\,\mu$m,
    $\rho=25\,$mm (c) $E=2.6\,\mathrm{GPa}$, $h=15\,\mu$m,
    $\rho=60\,$mm.  The parameter $\sigma$, defined in
    equation~(\ref{eq:delta2}), measures the strength of adhesion
    relative to the stiffness of the film.  (d) Sketch of a cut
    through a vertical plane $\mathrm{AA}'$.}
    \protect\label{fig:ExpPRL}
\end{figure}

In the present paper, we study some buckling patterns produced by
these experiments.  Specifically, we consider the case of the
capillary adhesion between a thin elastic film and a doubly-curved
substrate.  This geometry has been considered by one of us in a recent
experimental paper~\cite{hure2011}.  It built up on previous work
addressing the related case of a spherical shell adhering onto a
planar substrate~\cite{tamura0,tamura,springman2008}. In our 
experiments, a rigid
spherical cap is first coated by a wetting liquid and a thin
polypropylene film is then applied onto it.  As it wraps the sphere
under the action of capillary forces, the film is forced to stretch by
Gauss' \emph{theorema egregium}~\cite{struik}. Stretching allows it to
make up for the mismatch of Gaussian curvature, which is zero in the
planar film and non-zero along the spherical substrate.
This leads to a variety of
adhesion morphologies, as shown in figure~\ref{fig:ExpPRL}, where the contact region
varies from a simple band to complex branched patterns.
 
In reference~\cite{hure2011}, one of us studied the antagonistic
effects of elasticity and capillarity using order of magnitude arguments, and proposed an estimate
for the size of the region of adhesion which successfully compares to
the experiments.

Here, we study these patterns quantitatively.  In particular we
address the transition shown in the figure, whereby a band-like region
of contact with straight edges (figure~\ref{fig:ExpPRL}a) bifurcates
into a sinuous pattern with undulatory edges
(figure~\ref{fig:ExpPRL}b) when the adhesion becomes stronger or the
film becomes thinner.  This is interpreted as a buckling bifurcation
caused by compressive stress along the straight edges.
We carry out a stability analysis based on the classical Donnell
equations for nearly-cylindrical shells, modified to account for the
effect of adhesion.  The motion of the free boundary at the edge of
the region of contact is considered without any approximation.

The paper is organized as follows.  In section~\ref{bucklingproblem},
we derive the equations for a nearly-cylindrical elastic shell
adhering to a sphere, with an emphasis on the equilibrium conditions
along the edge of the moving contact region.  In
section~\ref{anonlinear}, we derive a non-linear solution to these
equations relevant to the unbuckled configuration with
cylindrical symmetry.  These results are compared to experimental
data.  In section~\ref{sectstab}, we then study the linear stability
of the cylindrical solution. The predictions regarding the symmetry of
the buckling modes, their wavelength and the critical loads are
compared to experimental data in section~\ref{sectcompexp}.

\section{Governing equations: Donnell's shell equations with adhesion}
\label{bucklingproblem}

The Donnell-Mushtari-Vlasov equations for nearly-cylindrical shells,
simply called the Donnell equations thereafter, are derived by
combining the general equations for shells undergoing finite
displacements with specific scaling assumptions for the magnitude of
the displacement.  Even though the Donnell equations have been known
and used for a long time, this derivation is useful as it highlights
the simple assumptions that underlie them.  More importantly, the
variational framework that is used to derive the Donnell equations
allows us to include adhesion and the presence of a moving boundary in
a natural and consistent manner.

\subsection{Geometry}

\begin{figure}
    \centering
    \includegraphics[width=.7\textwidth]{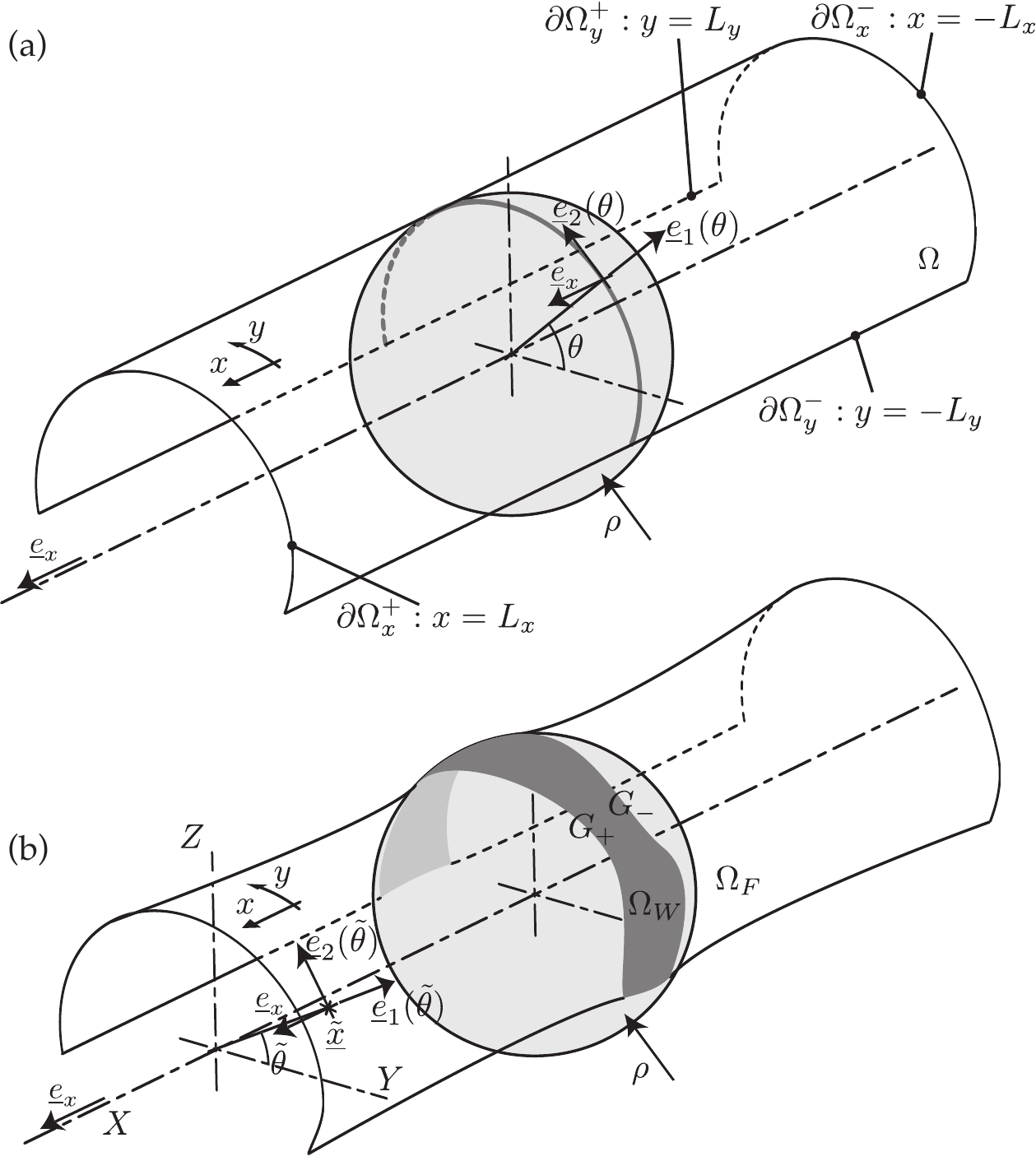}
    \caption{(a) The reference configuration: the cylindrical shell makes
    tangent contact with the sphere along the great circle $x=0$.  
    (b)~A
    typical deformed configuration: the region of contact
    $\Omega_{W}$, shown in dark grey, is bounded by two curves $G_{-}$
    and $G_{+}$.}
    \label{fig:ShellReferenceActual}
\end{figure}

The reference configuration considered here is shown in
figure~\ref{fig:ShellReferenceActual}a. 
A thin cylindrical shell $\Omega$ of half-length $L_{x}$, half-width
$L_{y}$, thickness $h$ and radius $\rho$ rests tangentially to a
sphere of radius $\rho$.  The Lagrangian coordinates along the shell
are denoted by $x$ and $y=\rho\,\theta$, where $\theta$ is the
azimuthal angle in the reference configuration.  Let $\partial
\Omega_x^{\pm}$ and $\partial \Omega_y^{\pm}$ correspond to the edges
of the shell at $x = \pm L_{x}$ and $y = \pm L_{y}$, respectively.  We
assume that the shell is made of an isotropic linear elastic material with Young's 
modulus $E$ and Poisson's ratio $\nu$.

A typical deformed configuration is shown in
figure~\ref{fig:ShellReferenceActual}b.  The displacement of the
middle surface of the shell is denoted $u_{x}(x,y)$, $u_{y}(x,y)$ and
$w(x,y)$, where the first two functions are the tangential components
of the displacement, and $w(x,y)$ is the radial displacement
(deflection), counted positive towards the exterior of the shell.  In
the deformed configuration, the position of a generic point
lying on the center-surface of the shell reads
\begin{subequations}
    \label{eq:deformedPos}
\begin{align}
    \underline{\tilde{x}}(x,y) & = \Big(x+u_{x}(x,y)\Big)\,\underline{e}_{x}
    + 
    \Big(
    \rho + w(x,y)
    \Big)\,
    \underline{e}_{1}(\tilde{\theta}(x,y))
    \textrm{,}
    \\
    \intertext{where the azimuthal angle in deformed configuration reads}
    \quad
    \tilde{\theta}(x,y) &=
    \frac{y+u_{y}(x,y)}{\rho}
    \textrm{.}
\end{align}
\end{subequations}
Here $\underline{e}_{1}(\tilde{\theta}) =
(0,\cos\tilde{\theta},\sin\tilde{\theta})$ is the radial unit vector
in the plane perpendicular to the axis $\underline{e}_{x}$.

The purely radial displacement $w_{s}\leq 0$ that brings a point from
the cylindrical reference configuration onto the sphere is such that
$x^2+(\rho+w_{s}(x))^2 = \rho^2$, and so
\begin{equation}
    w_{s} (x) = -(\rho - \sqrt{\rho^2 - x^2})
    \nonumber
\end{equation}
where $x$ is the axial coordinate and $r$ is the radial coordinate. 
In the following, we shall use a second-order approximation, valid 
for $|x|\ll\rho$,
\begin{equation}
    w_{s} (x) = -\frac{1}{2}\,\frac{x^2}{\rho}
    \textrm{.}
    \label{eq:sphereParabolicApproximDimensionalForm}
\end{equation}
In this approximation, the sphere has been effectively replaced by its
osculating paraboloid.  The non-penetration condition is then
expressed as a unilateral constraint,
\begin{equation}
    w(x)\geq w_{s}(x)
    \textrm{.}
    \label{eq:nonPenetration}
\end{equation}

We denote $\Omega_W$ the contact zone between the sphere and the
shell, and $\Omega_F$ the free part of the shell, $\Omega = \Omega_W
\cup \Omega_F$.  Neglecting the width of the meniscus, we assume that
the boundary $G = \Omega_W \cap \Omega_F$ between the two domains is
made up of two curves,
\begin{equation}
    G = G_{-}\cup G_{+}
    \textrm{.}
    \label{eq:GHasTwoComponents}
\end{equation}
In addition, we assume that each one of this curve is a graph,
\begin{align}
    G_{-} & = \{(x,y)\;|\; x = g_{-}(y)\} 
    \label{eq:G-}\\
    G_{+} & = \{(x,y)\;|\; x = g_{+}(y)\} 
    \textrm{.}
    \label{eq:G+}
\end{align}
This assumption is valid for configurations that are close to an
axisymmetric configuration, for which both the functions $g_{-}$ and
$g_{+}$ are constant.  The wet region  $\Omega_{W}$ lies inside the boundaries
$G_{-}$ and $G_{+}$ and the free region $\Omega_{F}$ lies outside,
\begin{align}
    \Omega_{W} & = \{(x,y)\;|\; g_{-}(x)\leq x \leq g_{+}(y)\}
    \label{eq:OmegaW}\\
    \Omega_{F} & = \{(x,y)\;|\; x \leq g_{-}(y)\textrm{ or } g_{+}(y)\leq
    x\}
    \textrm{.}
    \label{eq:OmegaF}
\end{align}
In the rest of the paper, Greek indices such as $\alpha$ or $\beta$
represent the Lagrangian coordinates, $\alpha = x$ or $\alpha = y$,
and follow the implicit summation convention for repeated indices.

\subsection{Scaling assumptions}

Before introducing the fundamental mechanical quantities, which are
the membrane strain $e_{\alpha\beta}$, the curvature strain
$b_{\alpha\beta}$, the membrane stress $n_{\alpha\beta}$ and the bending
moment $m_{\alpha\beta}$, we present the scaling assumptions that 
underlie the Donnell equations.

First, the thin-shell theory assumes that the following slenderness
parameter is small,
\begin{equation}
    \eta^2 = \frac{h}{\sqrt{12}\,\rho}
    \textrm{.}
    \label{eq:defEta}
\end{equation}
We shall therefore consider the limit
\begin{equation}
    \eta \ll 1
    \textrm{.}
    \label{eq:etaSmall}
\end{equation}

The Donnell equations assume that the shell remains close to the
cylindrical configuration of reference.  How close depends on the
small parameter $\eta$: as shown in \ref{app:DonnellJustification}, it
is natural to rescale the deflection by $\rho\,\eta^2$, in-plane
lengths by $\rho\,\eta$, and in-plane displacements by $\rho\,\eta^3$.
Therefore, we define rescaled lengths by
\begin{subequations}
    \label{eq:rescaling1}
    \begin{equation}
        \overline{x} = \frac{x}{\rho\,\eta}
	\qquad\qquad  \overline{y} =\frac{y}{\rho\,\eta},
        \label{eq:rescaling-XY}
    \end{equation}
    and the rescaled displacement by
    \begin{align}
        \overline{u}_{\alpha}(\overline{x},\overline{y}) & = 
	\frac{1}{\rho\,\eta^3}\,u_{\alpha}(x,y),
	\label{eq:rescaling-displa-U}\\
	\overline{w}(\overline{x},\overline{y})  &= 
	\frac{1}{\rho\,\eta^2}\,w(x,y)
	\textrm{.}
        \label{eq:rescaling-displa-W}
    \end{align}
\end{subequations}

The parabolic approximation of the sphere profile in
equation~(\ref{eq:sphereParabolicApproximDimensionalForm}) reads, in
dimensionless variables,
\begin{equation}
    \overline{w}_{\mathrm{s}}(\overline{x}) =
    -\frac{1}{2}\overline{x}^2
    \textrm{.}
    \label{eq:rescaling-displas}
\end{equation}
Note that the assumption under which this approximation has been
derived, $|x|\ll \rho$ is consistent with the new scales introduced
here: when $\overline{x}$ is of order 1 (and in fact, as long as it
remains smaller than the large number $\eta^{-1}$), then $|x|\ll \rho$
and the sphere is indeed well approximated by its osculating paraboloid.

For the sake of consistency, we rescale the membrane strain 
$e_{\alpha\beta}$
by $\eta^2$, the curvature strain
$b_{\alpha\beta}$
by $1/\rho$, the membrane stress
$n_{\alpha\beta}$
by $C\,\eta^2$ and the bending moment
$m_{\alpha\beta}$
by $B/\rho$, where $B=Eh^3/[12(1-\nu^2)]$ and $C=Eh/(1-\nu^2)$ denote
the bending and stretching moduli of the shell, respectively. This 
is written
\begin{subequations}
    \label{eq:rescaling2}
    \begin{align}
        \overline{e}_{\alpha\beta}(\overline{x},\overline{y})
	& = \frac{1}{\eta^2}\,e_{\alpha\beta}(x,y)\\
	\overline{b}_{\alpha\beta}(\overline{x},\overline{y}) &= 
	\rho\, b_{\alpha\beta}(x,y)
        \label{eq:rescaling-strain}\\
        \overline{n}_{\alpha\beta}(\overline{x},\overline{y}) & = 
	\frac{1}{C\,\eta^2}\,n_{\alpha\beta}(x,y) \\
	\overline{m}_{\alpha\beta}(\overline{x},\overline{y}) & =
	\frac{\rho}{B}\,m_{\alpha\beta}(x,y)
	\emph{.}
        \label{eq:rescaling-stress}
    \end{align}
\end{subequations}
The definition of the small parameter $\eta$ given earlier in
eq.~(\ref{eq:defEta}) warrants that the bending energy density
$\frac{1}{2}\,m_{\alpha\beta}\,b_{\alpha\beta}$ and the stretching
energy density $\frac{1}{2}\,n_{\alpha\beta}\,e_{\alpha\beta}$ are
commensurate when $\eta$ goes to zero.

\subsection{Membrane and curvature strains}

From now on, we shall use dimensionless quantities everywhere, and
drop bars to easy legibility.  The above scaling assumptions allow one
to simplify the expressions for the membrane and curvatures strains
from the general theory of shells as follows,
\begin{subequations}
    \label{eq:strainDonnell}
    \begin{align}
        e_{\alpha\beta}(x,y)  & = 
	\left(
	\frac{
	u_{\alpha,\beta}(x,y)
	+
	u_{\beta,\alpha}(x,y)
	}{2}
	+
	\delta_{\alpha y}\,\delta_{\beta y}\,w(x,y)
	\right)
	+\frac{1}{2}\,w_{,\alpha}(x,y)\,w_{,\beta}(x,y)
	\label{eq:DonnellMembraneStrain}\\
        b_{\alpha\beta}(x,y) & = -w_{,\alpha\beta}(x,y)
	\textrm{.}
        \label{eq:DonnellBendingStrain}
    \end{align} 
\end{subequations}
These expressions are shown to derive from the scaling assumptions in
\ref{app:DonnellJustification}.  Here we use the Kronecker delta
symbol, which is defined by
\begin{equation}
\delta_{\alpha \beta} =  
\begin{cases}
1 & \text{if } \alpha = \beta \\
0 & \text{if } \alpha \neq \beta 
\end{cases}
\label{eq:KroneckerDelta}
\end{equation}
and use commas in indices to denote partial derivatives, as in the 
expression
\begin{equation}
    u_{x,y}(x,y) = \frac{\partial u_{x}(x,y)}{
    \partial y
    }
    \textrm{.}
    \nonumber
\end{equation}

The strain approximation in equation~(\ref{eq:DonnellMembraneStrain})
is at the heart of Donnell's model for nearly-cylindrical shells, and has
been used by numerous authors, see for instance~\cite{amabili2008}.
Only the non-linear terms that are important near the onset of
buckling have been retained.

\subsection{Constitutive law and energy}

The shell's material is assumed to be linearly elastic and isotropic. The Hookean constitutive laws read:
\begin{subequations}
    \label{eq:constitutive}
\begin{align}
    n_{\alpha \beta}(x,y) &= (1 - \nu)\,e_{\alpha \beta}(x,y) +
    \nu\,\delta_{\alpha \beta} \,\big(\tr \underline{\underline{e}}(x,y)\big)
    \label{eq:constitutiveMembrane}
    \\
    m_{\alpha \beta}(x,y) & = (1 - \nu)\,b_{\alpha \beta}(x,y) +
    \nu\,\delta_{\alpha \beta} \,\big(\tr \underline{\underline{b}}(x,y)\big)
    \label{eq:constitutiveBending}
\end{align}
\end{subequations}
This is a rescaled form of the original constitutive laws given in the
Appendix in equation~(\ref{eq:formalExpansionConstitutive}).  Thanks
to our rescalings in equation~(\ref{eq:rescaling2}), the stretching
and bending moduli have effectively been set to one.

In the experiments, the fim is naturally planar and is bent into a
cylindrical shape by the action of capillary forces.  By contrast, the
constitutive equation~(\ref{eq:constitutiveBending}) describes a
\emph{naturally cylindrical} shell with natural radius $\rho$: the
bending moment $m_{\alpha\beta}$ is zero when the bending strain
$\underline{\underline{b}}$ cancels, which happens in the
configuration of reference.  The case of a naturally \emph{planar}
film could be treated by modifying bending strain, replacing
$\underline{\underline{b}}$ with $(\underline{\underline{b}} +
\kappa_{0}\,\underline{e}_{y}\otimes \underline{e}_{y})$ everywhere,
where $\kappa_{0} = -1/\rho$ is the curvature strain in reference
configuration.  As we shall see later, only the derivatives of the
bending moment enter into the local equations of equilibrium.  As a
result, $\kappa_{0}$ is absent from these equations, and only affects
boundary layers near the free edges.  To simplify the analysis of
these boundaries, we ignore this and set $\kappa_{0} = 0$.  This is a
valid approximation for the entire pattern, except very close to the
film boundaries.

The linear constitutive laws correspond to the following quadratic
elastic energy,
\begin{equation}
    E_{shell} = \iint_{\Omega}\frac{1}{2}
    \,(
    n_{\alpha\beta}\,e_{\alpha\beta}
    +
    m_{\alpha\beta}\,b_{\alpha\beta})
    \;\mathrm{d}x\,\mathrm{d}y
    \label{eq:DonnellEnergy}
\end{equation}
which is the sum of a stretching and a bending term.

The adhesion between the shell and the rigid sphere is modelled by an
energy $(2 \gamma)$ per unit area of the contact region.  Indeed the
liquid perfectly wets both the shell and the rigid sphere, and any
increase in the area of contact removes two air/liquid interfaces,
each having an energy $\gamma$ per unit area~\cite{Roman2010},
where $\gamma$ is the surface tension of the liquid.

The rescaled adhesion energy thus reads:
\begin{equation}
    E_{adhesion} = 
    -\iint_{\Omega} 2\, \sigma^2\,\raisedchi_W(G;x,y) \,\mathrm{d}x\, \mathrm{d}y
    \textrm{,}
\label{eq:energAdhesion}
\end{equation}
where $\raisedchi_W(G;x,y)$ is the characteristic function of the wet
domain: $\raisedchi_W(G;x,y)=1$ in the wet region $(x,y)\in \Omega_W$,
and $\raisedchi_W(G;x,y)=0$ in the free region.  The coefficient
$\sigma^{2}$ is the capillary energy rescaled by the typical energy
per unit area introduced earlier, $B/\rho^2$:
\begin{equation}
    \sigma^{2} = \frac{\gamma}{B/\rho^2} = \left(\frac{\rho}{L_{ec}} \right)^2
    \textrm{.}
    \label{eq:delta2}
\end{equation}
Here we have introduced the elastocapillary length~\cite{Roman2010},
\begin{equation}
    L_{ec}  =\sqrt{\frac{B}{\gamma}}  = 
    \sqrt{\frac{E\,h^3}{12\,(1-\nu^2)\,\gamma}}
    \textrm{,}
    \label{eq:Lec}
\end{equation}
which arises from a balance of the bending rigidity of the shell $B$
and the adhesion energy $\gamma$.  At length scales smaller than
$L_{ec}$, bending stiffness dominates capillary effects.  The
elastocapillary length sets the typical radius of the sphere beyond
which adhesion is possible, as noted in reference~\cite{hure2011}.

\subsection{Constraints and boundary conditions}

In the wet region, the condition of contact with the sphere reads:
\begin{equation}
    w(x,y) = w_{\mathrm{s}}(x)\quad \textrm{in }\Omega_{\mathrm{w}}
    \textrm{,}
    \label{eq:contactCondition}
\end{equation}
where $w_{\mathrm{s}}$ is the parabolic approximation of the
deflection of the rigid sphere derived in
equation~(\ref{eq:rescaling-displas}).  We assume that the contact is
frictionless.

We consider that the shell is infinitely long in its $y$ direction.
This is captured by the following boundary condition on the remote
edges $\partial\Omega_{y}^\pm$:
\begin{equation}
  u_y(x,L_{y}^\pm) = V^{\pm}
 \label{eq:edgeYcondition}
\end{equation}
where the two unknown scalars $V^{+}$ and $V^{-}$ denote an unknown
rigid-body motion of the edges consistent with the symmetry.
They will be determined later by a condition of equilibrium
of the boundary.  This boundary condition means that the remote edges
$\partial\Omega_{y}^\pm$ remain contained in a plane passing through
the $X$ axis,
even though this plane can freely rotate
about this axis to accommodate an average extension or contraction of
the shell in its $y$ direction.  In the experiments, the film has a
finite extent, and the above boundary condition is relevant to the
case where its size is much larger than the wavelength of the
instability: anticipating the notations of section~\ref{sectstab},
this writes $L_{y} \gg 2\pi/k_{\mathrm{c}}$.

Finally, we introduce a new set of variables $(q_{x},q_{y})$ which is
the local slope of the shell with respect to the mobile frame,
\begin{equation}
    q_{\alpha}(x,y) = w_{,\alpha} (x,y)
\label{eq:defW}
\end{equation}
In the following, these $q_{\alpha}$'s will be considered variables
independent from the deflection $w$, and the equation~(\ref{eq:defW})
just written will be viewed as a constraint.  This allows our
second-order variational problem to be written as a first-order one
having additional variables and constraints, and simplifies the
derivation of the equilibrium equations.

The Lagrangian of our constrained minimization problem is formed by
augmenting the elastic and adhesion energies in
equations~(\ref{eq:DonnellEnergy}) and (\ref{eq:energAdhesion}) with
the constraints in equations~(\ref{eq:contactCondition}),
(\ref{eq:edgeYcondition}) and~(\ref{eq:defW}) by means of Lagrange
multipliers, denoted $\pi(x,y)$, $\nu^{\pm}(x)$ and
$\lambda_{\alpha}(x,y)$:
\begin{multline}
    E_{tot}(u_{\alpha},w,q_{\alpha},V^\pm,G,\pi,\nu^{\pm},\lambda_{\alpha}) = 
\\
    \iint_{\Omega}\frac{1}{2}\,( 
    n_{\alpha\beta}\,e_{\alpha \beta} + 
    m_{\alpha \beta}\,b_{\alpha\beta} 
    )\,\mathrm{d}x\, \mathrm{d}y
    - 
    \iint_{\Omega} 2\,\sigma^2\,\raisedchi_W(G)\,\mathrm{d}x\, \mathrm{d}y
\\
    - \iint_{\Omega} \pi\,(w-w_{s})\,\raisedchi_W(G)\,\mathrm{d}x\, \mathrm{d}y
    -\sum_{s=\pm}\int_{\partial\Omega_{y}^s} \nu^{s}\,(u_y-V^s)\,\mathrm{d}x
    -\iint_{\Omega} \lambda_{\alpha}\,(w_{,\alpha} - q_{\alpha})\,\mathrm{d}x\, \mathrm{d}y
    \textrm{.}
\label{eq:energy}
\end{multline}
The interpretation of the Lagrange multipliers just introduced will be
given in section~\ref{eqequi}.  Note that the
constraints~(\ref{eq:edgeYcondition}) on the boundaries $\partial
\Omega_{y}^-$ and $\partial \Omega_{y}^+$ are taken care of using a
summation over the values $s=-$ and $s=+$.

\subsection{Equations of equilibrium}
\label{eqequi}

In this section, we derive the equilibrium equations and the boundary
conditions using variational calculus, by cancelling the first
variation of the Lagrangian just written.  This yields the classical
Donnell equations for shells inside each domain $\Omega_W$ and
$\Omega_F$, as well as boundary conditions, including an adhesion
condition at the interface $G$ between the wet and free regions.

We first compute the variation of the Lagrangian with respect to the
unknowns
$(u_{\alpha},w,q_{\alpha},V^\pm,\pi,\nu^{\pm},\lambda_{\alpha})$
considering that domains $\Omega_W$ and $\Omega_F$ and their boundary
$G$ remain fixed, as expressed by the notation $\delta g=0$,
\begin{multline}
    \delta E_{tot}^{\dag} = 
    \delta E_{tot}(
    \delta u_{\alpha},\delta w,\delta q_{\alpha},\delta V^\pm,\delta g_{\pm} = 0 , \delta \pi,\delta \nu^{\pm},\delta \lambda_{\alpha})
    = \\
    \iint_{\Omega}
    (
    n_{\alpha \beta}\, \delta e_{\alpha \beta} 
    +
    m_{\alpha \beta} \,\delta b_{\alpha\beta} 
    )
    \;\mathrm{d}x\, \mathrm{d}y
    - \iint_{\Omega} \pi \,\raisedchi_W(G)\,\delta w
    \;\mathrm{d}x\, \mathrm{d}y
    \\
    - \sum_{s=\pm}
    \int_{\partial\Omega_{y}^s} \,\nu^{s} \, \delta u_y
    \;\mathrm{d}x
    + \sum_{s=\pm}
    \left(\int_{\partial \Omega_{y}^{s}}\nu^s \;\mathrm{d}x\right)\,\delta V^s
   - \iint_{\Omega} \lambda_{\alpha}\,(\delta w_{,\alpha} - \delta q_{\alpha})\,\mathrm{d}x\, \mathrm{d}y
   \\
    - \iint_{\Omega}(w-w_{s})\,\raisedchi_W(G)\, \delta \pi
    \;\mathrm{d}x\, \mathrm{d}y
    -\sum_{s=\pm}\int_{\partial\Omega_{y}^s}
    (u_y-V^s)\, \delta \nu^{s}
    \;\mathrm{d}x
    -\iint_{\Omega}(w_{,\alpha} - 
    q_{\alpha})\, \delta \lambda_{\alpha}
    \;\mathrm{d}x\, \mathrm{d}y
\end{multline}
where we use the fact that $\delta(n_{\alpha \beta}\,e_{\alpha \beta})
= 2\,n_{\alpha \beta} \,\delta e_{\alpha \beta} $ and $\delta(m_{\alpha
\beta}\,b_{\alpha \beta}) = 2\,m_{\alpha \beta}\, \delta b_{\alpha
\beta}$, as we assume linear constitutive laws.  The three last
terms, which come from the variation with respect to the Lagrange
multipliers $(\pi,\nu^{\pm},\lambda_{\alpha})$, enforce the kinematic
relations in equations~(\ref{eq:contactCondition}), (\ref{eq:edgeYcondition}),
(\ref{eq:defW}), as expected.

The complementary variations with respect to the position of the
boundary $G$ will be computed later, with the help of
\ref{app:CornerConditions}.  They yield jump conditions across the
boundary, which include the condition of adhesion.

Calculating the first variation of the strain defined in
equation~(\ref{eq:strainDonnell}), and using the symmetry of the
stress tensors $n_{\alpha\beta}$ and $m_{\alpha\beta}$, we have $
n_{\alpha \beta}\,\delta e_{\alpha \beta} = n_{\alpha \beta}\,(\delta
u_{\alpha,\beta} + \delta_{\alpha y}\,\delta_{\beta y}\,\delta w +
w_{\beta}\,\delta w_{,\alpha})$ and $m_{\alpha\beta}\, \delta
b_{\alpha \beta} = -m_{\alpha\beta}\,\delta q_{\alpha,\beta}$.  Note
that the symbol $\delta$ without any subscript denotes the variation
of a function, while $\delta_{\alpha \beta}$ with subscripts is the
Kronecker delta symbol introduced in
equation~(\ref{eq:KroneckerDelta}).  Integrating by parts and grouping
the terms, we have:
\begin{multline}
    \delta E_{tot}^{\dag}  = 
    -
    \iint_{\Omega} 
    \Big(
    \big[
    n_{\alpha\beta , \beta}
    \big] \,\delta u_{\alpha} 
    + 
    \big[
    \pi\,\raisedchi_W(G) + (n_{\alpha\beta}\, 
    w_{,\alpha})_{,\beta} -n_{yy}
    -\lambda_{\alpha,\alpha}
    \big]\,\delta w 
    +
    \big[-m_{\alpha\beta,\beta}-\lambda_{\alpha}
    \big]\,\delta q_{\alpha}
    \Big)\,\mathrm{d}x\, \mathrm{d}y \\
    \\
    +\int_{\partial \Omega}
    \Big(
    \big[
    n_{\alpha \beta}
    -\delta_{\alpha y}\,\sum_{s=\pm}\nu^{s}\,
    \raisedchi_{\partial 
    \Omega_{y}^{s}}\,
    \big]\,\delta u_{\alpha} 
    + 
    \big[
    n_{\alpha \beta} \,w_{,\alpha} - \lambda_{\beta}
    \big]\,\delta w 
    -
    \big[m_{\alpha \beta}\big]\,\delta q_{\alpha}
    \Big)\,N_{\beta}\,\mathrm{d}\ell
    + \sum_{s=\pm}\left(
    \int_{\partial \Omega_{y}^{s}}\nu^{s}
    \,\mathrm{d}x\right)\,\delta V^s
\label{eq:deltaEintparts}
\end{multline}
where we denote the entire lateral boundary by $\partial\Omega =
\partial\Omega_{x}^\pm \cup \partial\Omega_{y}^\pm$, the element of
length along the lateral boundary by $\mathrm{d}\ell = \mathrm{d}x$ on
$\partial\Omega_{y}^\pm$ and $\mathrm{d}\ell = \mathrm{d}y$ on
$\partial\Omega_{x}^\pm$.  In addition, $\underline{N}=(N_x,N_y)$
stands for the normal to a boundary $\partial \Omega$, $\underline{N}$
being oriented towards the exterior of the domain $\Omega$. 

The equilibrium condition, $\delta E_{tot}^{\dag} = 0$, yields the
following equations on $\Omega$:
\begin{subequations}
\label{eq:equilibrium}
    \begin{align}
    n_{\alpha \beta , \beta}(x,y) &= 0
    \label{eq:shellEqLongitudinal}\\
    m_{\alpha \beta, \alpha\beta}(x,y)
    +n_{\alpha \beta}(x,y)\, w_{,\alpha \beta}(x,y)  - n_{yy}(x,y) + 
    \raisedchi_W(G;x,y) \, \pi(x,y)
    \label{eq:shellEqTransverse}
    &= 0
    \textrm{.}
    \end{align}
\end{subequations}
Here we have replaced the shear force $\lambda_{\alpha}$ by its
expression $\lambda_{\alpha} = -m_{\alpha \beta,\beta}$.  The latter
comes from the condition associated with perturbations $\delta
q_{\alpha}$.

The equations~(\ref{eq:equilibrium}) are the Donnell equations for
thin cylindrical shells, see for instance~\cite{amabili2008}.  The
first equation~(\ref{eq:shellEqLongitudinal}) is a tangential balance
of forces.  The second equation~(\ref{eq:shellEqTransverse}) is a
transverse balance of force.  The Lagrange multiplier $\pi$ can be
interpreted as the pressure of contact with the sphere in the wet
region.  The equation~(\ref{eq:shellEqTransverse}) allows one to
compute the contact pressure $\pi(x,y)$ in the wet region where
$w=w_{s}$ is known; in the free region, the pressure term $\pi(x,y)$
is zero and the equation is an equation for the unknown deflection
$w(x,y)$.

On the lateral boundary $\partial \Omega_x^\pm$, we recover the
natural boundary conditions for a plate or shell with a free edge, see
for instance~\cite{mansfield89},
\begin{subequations}
\label{eq:boud}
\begin{align}
n_{\alpha x}(\pm L_{x},y) & = 0\label{eq:boud-n}\\
m_{xx}(\pm L_{x},y) & = 0\\
m_{xx,x}(\pm L_{x},y)+ 2\,m_{xy,y}(\pm L_{x},y)& = 0
\textrm{.}
\label{eq:boudshear}
\end{align}
\end{subequations}
Note that equation~(\ref{eq:boudshear}) comes from the
integration by parts of $\delta q_y$ in equation~(\ref{eq:deltaEintparts}),
as $\delta q_y = (\delta w)_{,y}$ on $\partial \Omega_{x}^\pm$.

On the other lateral boundaries $\partial\Omega_{y}^\pm$, the boundary
conditions read:
\begin{subequations}
\label{eq:boundnu0}
\begin{align}
n_{xy}(x,\pm L_{y}) & = 0\\ 
n_{yy}(x,\pm L_{y}) & = \nu^\pm(x) \\
m_{yy}(x,\pm L_{y}) & = 0\\
m_{yy,y}(x,\pm L_{y})+ 2\,m_{xy,x}(x,\pm L_{y}) &= -\nu^\pm(x) w_{,y}(x,\pm L_{y}) \\
\int_{\partial\Omega_{y}^\pm} n_{yy}(x,\pm L_{y})\,\mathrm{d}x  &= 0 
\textrm{.}
\label{eq:boundnu}
\end{align}
\end{subequations}
Here we used again the fact $\delta q_x = (\delta w)_{,x}$ on
$\partial \Omega_{y}^\pm$.  The Lagrange multiplier $\nu^{\pm}$ can be
interpreted as the normal stress on $\partial\Omega_{y}^\pm$.  The
equation~(\ref{eq:boundnu}) expresses the fact that no average force
is applied on the shell in the $y$ direction (natural boundary
condition).

\subsection{Equations for the moving boundary}

We recall the general expression of the corner conditions in a
two-dimension domain in \ref{app:CornerConditions}, known as the
Weierstrass-Erdmann conditions.  They yield the jump conditions at the
moving interface between two subdomains, in any minimization problem
where the contributions to the objective function have different
expressions in each subdomain.  This applies to the Lagrangian of our
problem in equation~(\ref{eq:energy}), which can indeed be written as
\begin{equation}
    E_{tot} = \iint_{\Omega_W} \mathcal{L}^W(u_{\alpha},w,q_{\alpha},\pi,\lambda_{\alpha})\,\mathrm{d}x\, \mathrm{d}y
+ \iint_{\Omega_F} \mathcal{L}^F(u_{\alpha},w,q_{\alpha},\pi,\lambda_{\alpha})\,\mathrm{d}x\, \mathrm{d}y
+ \int_{\partial \Omega} \cdots \,\mathrm{d}\ell
\textrm{.}
\label{eq:energErdmann}
\end{equation}
The integrands have different expressions in the free and wet
regions,
\begin{align}
    \mathcal{L}^{F}(u_{\alpha},w,q_{\alpha},\pi,\lambda_{\alpha};x,y) & 
    = \frac{1}{2}\,
    ( 
    n_{\alpha\beta}\,e_{\alpha \beta} + 
    m_{\alpha \beta}\,b_{\alpha\beta} 
    )- \lambda_{\alpha}\,(w_{,\alpha} - q_{\alpha}) \\ 
    \mathcal{L}^{W}(u_{\alpha},w,q_{\alpha},\pi,\lambda_{\alpha};x,y) & 
    = \mathcal{L}^{F}(u_{\alpha},w,q_{\alpha},\pi,\lambda_{\alpha})
    -  2\, \sigma^2
    -  \pi\,(w-w_{\mathrm{s}})
    \emph{.}
\end{align}
These notations conform with those of equation~(\ref{eq:erdmanenergy})
in \ref{app:CornerConditions}.  As usual with shell models having
non-zero bending rigidity, the tangent displacement $u_{\alpha}(x,y)$
is required to be continuous, and the transverse displacement $w(x,y)$
to be $\mathcal{C}^1$-smooth.  In particular, across the boundary $G$,
we have
\begin{subequations}
    \label{eq:discont-1}
    \begin{align}
        \discont{u_{\alpha}} & = 0 \\
	\discont{w} & = 0 \\
	\discont{w_{,\alpha}} & = 0 
	\textrm{,}
    \end{align}
\end{subequations}
where 
\begin{equation}
    \discont{A} = A^{F}(x_{G},y_{G}) - A^{W}(x_{G},y_{G})
    \label{eq:jumpDoubleBracket}
\end{equation}
denotes the discontinuity of a function $A$ across a point
$(x_{G},y_{G})$ lying on the boundary $G$.  

By differentiation of the equalities~(\ref{eq:discont-1}) along the
boundary $G$, we find
\begin{subequations}
    \label{eq:discont-2}
    \begin{align}
        \discont{u_{\alpha,T}} & = 0 \\
        \discont{w_{,\alpha T}} & = 0
	\textrm{,}
    \end{align}
\end{subequations}
where $\underline{T}=(T_x,T_y)$ is the unit tangent to $G$, and a $T$
in subscript denotes the tangent derivative $f_{,T} =
T_{\alpha}\,f_{,\alpha}$.  The functions $w$ and $q_{\alpha} =
w_{,\alpha}$ have prescribed values in the wet region $\Omega_W$ but
can take any value in the free region $\Omega_F$.  By contrast,
$u_{\alpha}$ is unconstrained in the entire domain $\Omega$.

Using the fact that the elastic energy is a quadratic form of the
strain by equation~(\ref{eq:DonnellEnergy}) first, and using the
definition of the strain in equation~(\ref{eq:strainDonnell}) next,
one can compute the so-called generalized momentum and identify the
result with the internal stress, up to a sign,
\begin{alignat}{3}
    \frac{\partial \mathcal{L}^i}{\partial u_{\alpha,\beta}}
    & = n_{\gamma\rho}^i\,\frac{\partial e_{\gamma\rho}}{\partial u_{\alpha,\beta}}
    & & = n_{\gamma\rho}^i\,\frac{1}{2}\,(
    \delta_{\alpha\gamma}\,\delta_{\beta\rho}
    +
    \delta_{\alpha\rho}\,\delta_{\beta\gamma}
    )
    & & =
    n_{\alpha\beta}^i
    \label{eq:momentumIdentificationN}
    \\
    \frac{\partial \mathcal{L}^i}{\partial q_{\alpha,\beta}} 
    & =  m_{\gamma\rho}^i
    \,\frac{\partial b_{\gamma\rho}}{\partial q_{\alpha,\beta}}
    & & = 
    -
    m_{\gamma\rho}^i\,
    \frac{1}{2}\,(
    \delta_{\alpha\gamma}\,\delta_{\beta\rho}
    +
    \delta_{\alpha\rho}\,\delta_{\beta\gamma}
    )
    & & = - m_{\alpha\beta}^i
    \label{eq:eq:momentumIdentificationM}
\end{alignat}
where $i=W,F$ is any of the wet (W) or free (F) region. 

We now apply the corner conditions derived
in~\ref{app:CornerConditions} after identifying the regions $\Omega_1
=\Omega_W $, $\Omega_2 =\Omega_F $.  The unknowns, collectively denoted
$\xi_{\alpha}$  in the appendix, are the in-plane
displacement $u_{\beta}$, the deflection $w$ and the slope
$q_{\beta}$.

When the equilibrium condition~(\ref{eq:eqjj1}) derived in the
appendix is applied to the unknown $\xi_{\alpha } = u_{\alpha}$, we
find $\discont{n_{\alpha\beta}}\, N_{\beta} = \discont{n_{\alpha N}} =
0$ after using equation~(\ref{eq:momentumIdentificationN}).  Note that
$N_{\beta}$ refers to the local normal vector, while $n_{\alpha\beta}$
refers to a generic component of the membrane stress.  For $\alpha =
T$ and $\alpha = N$ this yields $\discont{n_{TN}} = 0$ and
$\discont{n_{NN}} = 0$, which are the classical conditions for the
in-plane equilibrium of the boundary.

It turns out that the third independent component $n_{TT}$ of the
membrane stress is also continuous across the boundary, even though
this does not directly follow from equilibrium.  To show this,
let us write the constitutive law~(\ref{eq:constitutiveMembrane}) in
dimensionless form, $n_{TT} = \nu\,n_{NN} + (1-\nu^2)\,e_{TT}$.  In
the right-hand side, $n_{NN}$ is continuous as we have just shown,
while $e_{TT}$ is continuous as a consequence of the smoothness
conditions~(\ref{eq:discont-1}) and~(\ref{eq:discont-2}).  Therefore,
the tangential stress $n_{TT}$ is continuous as well.
To sum up, we have shown that all components of the membrane stress
are continuous,
\begin{equation}
    \discont{n_{\alpha\beta}} = 0
    \textrm{.}
    \label{eq:membraneStressContinuity-tmp-TT}
\end{equation}
Using the constitutive law for stretching in its inverted form, one
shows that the membrane strain is continuous as well,
\begin{equation}
    \discont{e_{\alpha \beta}}  = 0
    \textrm{.}
\end{equation}
As a result, the density of stretching energy is continuous,
$\discont{\frac{1}{2}\,n_{\alpha \beta}\,e_{\alpha \beta}} = 0$.  

The
discontinuity in the energy density is therefore the sum of an adhesion
term, which is present only in the wet part, and a discontinuity in
bending energy,
\begin{equation}
    \discont{\mathcal{L}} = \mathcal{L}^F - \mathcal{L}^W
    = 2\,\sigma^2 + 
    \frac{1}{2}\,\discont{m_{\alpha\beta}\,b_{\alpha\beta}}
\textrm{.}
\end{equation}

Considering now the condition~(\ref{eq:eqjj}) for equilibrium of the 
boundary with respect to the variable $\xi_{\alpha} = q_{\alpha} = 
w_{,\alpha}$, we have
\begin{equation}
    2\,\sigma^2 + 
    \frac{1}{2}\,\discont{m_{\alpha\beta}\,b_{\alpha\beta}}
    - m_{\gamma 
    \beta}^F\,\discont{b_{\gamma\beta'}}\,N_{\beta}\,N_{\beta'}
    =0
    \textrm{.}
    \label{eq:equilBoundary-v1}
\end{equation}
It is consistent to use this equation~(\ref{eq:eqjj}) as the value of
$q_{\alpha} = w_{,\alpha} = w_{s,\alpha}$ is prescribed in the
adhering region.

Upon insertion of the constitutive law first, and of the continuity
relations for the curvature next, one can rewrite the second term of
equation~(\ref{eq:equilBoundary-v1}) as
\begin{equation}
    \frac{1}{2}\,\discont{m_{\alpha\beta}\,b_{\alpha\beta}}
    =
    \frac{1}{2}\,\discont{b_{NN}^2 + b_{TT}^2 + 
    2\,\nu\,b_{TT}\,b_{NN}+2\,(1-\nu)\,b_{TN}^2} \\
    =
    \frac{1}{2}\,\left(
    \discont{b_{NN}^2}+2\,\nu\,b_{TT}\,\discont{b_{NN}}
    \right)
    \textrm{.}
    \label{eq:equilBoundary-int1}
\end{equation}
Note that only the $NN$-component of $b_{\alpha\beta} =
-w_{,\alpha\beta}$ is possibly discontinuous.  As a result, only the
set of indices $\gamma = N$ and $\beta' = N$ need be considered in the
last term of~(\ref{eq:equilBoundary-v1}).  Using the constitutive law
for bending~(\ref{eq:constitutiveBending}) again, we can rewrite this
last term as follows,
\begin{equation}
    - m_{\gamma 
    \beta}^F\,\discont{b_{\gamma\beta'}}\,N_{\beta}\,N_{\beta'} = 
    -(b_{NN}^F+\nu \,b_{TT})\,\discont{b_{NN}}
    \textrm{.}
    \label{eq:equilBoundary-int2}
\end{equation}

Inserting the expressions~(\ref{eq:equilBoundary-int1})
and~(\ref{eq:equilBoundary-int2}) into the equilibrium
condition~(\ref{eq:equilBoundary-v1}), we find that the terms
proportional to Poisson's ratio cancel out:
\begin{equation}
    2\,\sigma^2 +
     \frac{1}{2}\,
    \left((b_{NN}^F)^2 - (b_{NN}^W)^2\right)
    -b_{NN}^F\,\left(b_{NN}^F - b_{NN}^W\right)
    =0
    \textrm{.}
\end{equation}
After using the definition of the jump operator in
equation~(\ref{eq:jumpDoubleBracket}), this can be simplified as
\begin{equation}
\label{eq:jumpcarre}
    2\,\sigma^2 -\frac{1}{2}\,\discont{b_{NN}}^2 = 0
    \textrm{.}
\end{equation}
By developing in the non-penetration condition in
equation~(\ref{eq:nonPenetration}) in Taylor series up to second
order, and by using the continuity of the deflection and the slope in
equations~(\ref{eq:discont-1}), we have
\begin{equation}
    w_{,NN}(x_G,y_G) \geq w_{s,NN}(x_G,y_G)
    \textrm{.}
\end{equation}
In terms of curvature, this yields $b_{NN}^F(x_G,y_G) \leq 
b_{NN}^W(x_G,y_G)$, which we rewrite using the jump notation as
\begin{equation}
 \discont{b_{NN}} \leq 0
 \textrm{.}
\end{equation}
Thus, the equation (\ref{eq:jumpcarre}) implies
\begin{equation}
\label{eq:jumpsanscarre}
    \discont{b_{NN}} = -2\,\sigma
    \textrm{,}
\end{equation}
as $\sigma$ is a positive number.

To sum up, there are seven independent continuity and jump relations
that must be enforced at the moving interface $g$. They read
\begin{subequations}
\label{eq:eqcontinuity}
\begin{align}
    \discont{u_{x}}_{x=g(y)} &= 0 \\
    \discont{u_{y}}_{x=g(y)} &= 0 \\
    \discont{w}_{x=g(y)} &= 0 \label{eq:continuity-W}\\
    \discont{w_{,N}}_{x=g(y)} &= 0 \label{eq:continuity-Wn} \\
    \discont{n_{T N}}_{x=g(y)} &= 0 \\
    \discont{n_{N N}}_{x=g(y)} &= 0 \\
    \discont{m_{NN}}_{x=g(y)}  &= -2\, \sigma \label{eq:mNNJump}
    \textrm{.}
\end{align}
\end{subequations}
Equation~(\ref{eq:jumpsanscarre}) comes from
equation~(\ref{eq:mNNJump}) after using the constitutive law.  The
notation $x=g(y)$ emphasises the fact that the jump operator is
defined on the boundary $G$ whose equation is $x=g_{\pm}(y)$.  A term
proportional to the perturbation $\delta g_{\pm}$ of the boundary will
appear in these equations when we consider the linear stability later
on.  This perturbation $\delta g_{\pm}$ must be considered as the
shape of the boundary may be affected by the buckling.  It corresponds
to the quantity $a_{1}$ introduced later.

The jump relation~(\ref{eq:mNNJump}) arising from adhesion has
previously been derived in an effectively 1D (axisymmetric)
case~\cite{seifert91,majidi09}, as well as for elastic
plates~\cite{gioia,majidi10}.  This equation~(\ref{eq:mNNJump}) is
also known in the context of classical fracture mechanics, as the
delamination of a thin film on a rigid substrate can be seen as the
propagation of a crack along an interface~\cite{suo}: the energy
release rate $G$ per unit width of a film adhering on a curved
substrate reads $G = \frac{\discont{m_{NN}}^2}{2B}$, and
equation~(\ref{eq:mNNJump}) expresses from the balance of energy,
$G=2\,\gamma$.

\section{A non-linear solution for the unbuckled state}
\label{anonlinear}

In this section we derive an axisymmetric solution to the equations
derived in section~\ref{bucklingproblem}.  The wet region $\Omega_{W}$
is then a strip bounded by two symmetric circles with equation $x =
g_{\pm}(y)$, where $g_{\pm}(y) = \pm a_{0}$.  Here $a_{0}$ denotes
half the width of the strip, and will determined later as a function
of the adhesion number $\sigma$.

In the axisymmetric case, the displacement is of the form
\begin{subequations}
    \label{eq:axisymDisplacement}
\begin{align}
    u_{x}(x,y) &= u_{x}^{0i}(x)
    \label{eq:axisymDisplacement-Ux}\\
    u_{y}(x,y) &= E_{0}\,y
    \label{eq:axisymDisplacement-Uy} \\
    w(x,y) & = w_{0}^i(x)
    \textrm{,}
\end{align}
\end{subequations}
where the index $i$ denotes the region ($i=W$ if $|x|\leq a_{0}$ and
$i=F$ if $|x|\geq a_{0}$), the symbol `$0$' refers to the unbuckled
axisymmetric state.  The unknown constant $E_{0}$ measures the uniform
hoop strain, and will be determined later.  It is related to the
unknown uniform tangent displacement $V^\pm$ at the edges $y = \pm
L_y$ by $E_{0} = u_{y,y} = (V^+-V^-)/(2\,L_{y})$.  Imposing a linear
dependence of $u_{y}$ on the azimuthal variable $y$ warrants that the
hoop strain will be independent of $y$, as required by the symmetry.

We compute the membrane strains
\begin{subequations}
    \label{eq:strainAxisym}
    \begin{align}
        e_{xx}^0(x,y) & = {u_{x}^{0i}}'(x) + 
	\frac{1}{2}\,({w_{0}^i}'(x))^2
        \label{eq:strainAxisym-xx}\\
        e_{xy}^0(x,y) & =0
        \label{eq:strainAxisym-xy}\\
        e_{yy}^0(x,y) & = E_{0} + w_{0}^i(x)
	\textrm{.}
        \label{eq:strainAxisym-yy}
    \end{align}
They are all independent of the coordinate $y$, as required by the
symmetry.  All bending strains are zero, except for the axial
component
\begin{equation}
    \underline{\underline{b}}^0(x,y) = b^0_{xx}\,
    \underline{e}_{x}\otimes \underline{e}_{x},
    \quad
    b^0_{xx}  = -{w_0^i}''(x)
    \textrm{.}
    \label{eq:bendingStrainAxisym}
\end{equation}
\end{subequations}

By the constitutive law for stretching we have $n_{xy}(x,y)=0$ and
$n_{\alpha \beta,y}(x,y) = 0$, as required by the symmetry.  The
equilibrium condition~(\ref{eq:shellEqLongitudinal}) along the
direction $x$ ($\alpha = x$ in this equation) implies that the stress
$n_{xx}$ does not depend on $x$ either.  By the boundary
condition~(\ref{eq:boud-n}), this quantity vanishes everywhere,
$n_{xx}^0(x,y) = 0$.  In terms of the in-plane strain, this writes
$e_{xx} + \nu\,e_{yy} = 0$, an equation which will be
useful to reconstruct the in-plane displacement $u_{x}^{0i}(x)$:
\begin{equation}
    {u_{x}^{0i}}'(x) = -\frac{1}{2}({w_{0}^i}'(x))^2
    -\nu\,w_{0}^i(x)
     - \nu\,E_{0}
     \textrm{.}
    \label{eq:axisymmReconstructUx}
\end{equation}
Elimination of $e_{xx} = -\nu\,e_{yy}$ from the
constitutive relation~(\ref{eq:constitutiveMembrane}) yields the
expression of the only non-zero stress component,
\begin{equation}
    \underline{\underline{n}}^{0}(x,y) = n_{yy}^0\,
    \underline{e}_{y}\otimes \underline{e}_{y}
    ,\qquad
    n_{yy}^0 = (1-\nu^2)\,e^0_{yy}(x,y)
    = (1-\nu^2)\,(E_{0} + w_{0}^i(x))
    \textrm{.}
    \label{eq:prestretch}
\end{equation}

In the wet region $|x|\leq a_{0}$,  the contact condition reads
\begin{equation}
    w^W_{0}(x) = w_{s}(x) = -\frac{1}{2}\,x^2
    \textrm{.}
    \label{eq:wWAxisym}
\end{equation}
The membrane stress there is then found by inserting this expression
into equation~(\ref{eq:prestretch}), up to the constant $E_{0}$ that
will be determined later.  The in-plane displacement ${u_{x}^{0W}}(x)$
is found by integration of equation~(\ref{eq:axisymmReconstructUx})
with the initial condition ${u_{x}^{0W}}(0)=0$ imposed by the symmetry.

By symmetry, many term cancel in the
equation~(\ref{eq:shellEqTransverse}) for the transverse equilibrium
in the free region $|x|\geq a_0$, which reads
\begin{equation}
    \frac{1}{1-\nu^2}\,{w_{0}^F}''''(x) +  w_{0}^F(x) = -E_{0}
    \label{eq:axisymFreeEquilibrium}
    \textrm{.}
\end{equation}

We consider the case of a shell of infinite length in the
$x$-direction, $L_{x} \to \infty$.  This corresponds to the
situation in the experiments where the length $L_{x}$ of the film is
much larger than the width of the wet region, that is much larger than
the typical in-plane length $\sqrt{\rho\,h}$ which we used to make
lengths dimensionless.  The generic solution of equation~(\ref{eq:axisymFreeEquilibrium})
that is bounded near $x\to \pm\infty$ reads
\begin{equation}
    w^F_{0}(x) = - E_{0} 
    + e^{\pm \frac{x}{x^*(\nu)}}\,\left(
    A_{1}\, \cos \frac{x}{x^*(\nu)}
    + A_2 \, \sin \frac{x}{x^*(\nu)}
    \right)
    \label{eq:wFAxisym}
\end{equation}
where $A_{1}$ and $A_{2}$ are unknown amplitudes, and
\begin{equation}
    x^*(\nu) = \left(
    \frac{4}{1-\nu^2}
    \right)^{1/4}
    \label{eq:defXStar}
\end{equation}
is a scaling factor applicable to in-plane lengths.  This $x^*(\nu)$
is a known function of Poisson's ratio.

In order to determine the four constants $(A_1,A_2,E_{0},a_{0})$, we
use all the boundary conditions that are not automatically satisfied,
namely the continuity conditions~(\ref{eq:continuity-W})
and~(\ref{eq:continuity-Wn}), the equilibrium condition for the
average force along the $y$-direction~(\ref{eq:boundnu}), and the jump
condition~(\ref{eq:mNNJump}) depending on the adhesion number
$\sigma$:
\begin{subequations}
    \label{eq:axisymMissingConds}
\begin{align}
w_{0}^F(a_{0}) &= w_{0}^W(a_{0})\\
{w_{0}^F}'(a_{0}) & = {w_{0}^W}'(a_{0})\\
\int_0^{+\infty} n^0_{yy}(x)\,\mathrm{d}x & = 0\\
m_{xx}^F(a_{0}) & = m_{xx}^W(a_{0}) -  2 \,\sigma
\end{align}
\end{subequations}
Even though some of the equations for the problem are non-linear,
equations~(\ref{eq:axisymMissingConds}) happen to be linear with
respect to the unknowns $A_1$, $A_2$ and $E_{0}$, when we make use of
equations~(\ref{eq:prestretch}), (\ref{eq:wWAxisym})
and~(\ref{eq:wFAxisym}).  This leads to a set of linear equations whose
coefficients depend non-linearly on their arguments,
\begin{subequations}
    \label{eq:sysl}
    \begin{equation}
	\underline{\underline{M}}\left(\frac{a_{0}}{x^*(\nu)},\sigma\right)\cdot
	\underline{X}
	= \underline{0}
	\label{eq:sysl-Syst}
    \end{equation}
    where
    \begin{equation}
	\underline{X} = 
	\Big(
	A_{1},A_{2},E_{0},{a_{0}}^2
	\Big)
	\label{eq:sysl-Unknown}
\end{equation}
and
\begin{equation}
    \underline{\underline{M}}
    \left(\hat{a} = \frac{a_{0}}{x^*(\nu)}, \sigma\right) =
    \begin{pmatrix}
	  e^{-\hat{a}} \,\cos \hat{a}
	& e^{-\hat{a}}\, \sin \hat{a} 
	& -1
	& \frac{1}{2}\\
	- e^{-\hat{a}}\,(\cos \hat{a} +  \sin \hat{a}) 
	& e^{-\hat{a}}\,(\cos \hat{a} -  \sin \hat{a}) 
	& 0
	& \frac{1}{\hat{a}}\\
	  e^{-\hat{a}}\,(\cos \hat{a} -  \sin \hat{a}) 
	& e^{-\hat{a}}\,(\cos \hat{a} +  \sin \hat{a})  
	& 0
	& -\frac{\hat{a}}{3} \\
	e^{-\hat{a}}\, \sin \hat{a} &
	-e^{-\hat{a}}\, \cos \hat{a} 
	& 0
	& \frac{1}{\hat{a}^2}\,\left(\frac{1}{2} + \sigma\right)
\end{pmatrix}
\textrm{.}
\end{equation}
\end{subequations}

\begin{figure}
    \centerline{\includegraphics[width=.8\textwidth]{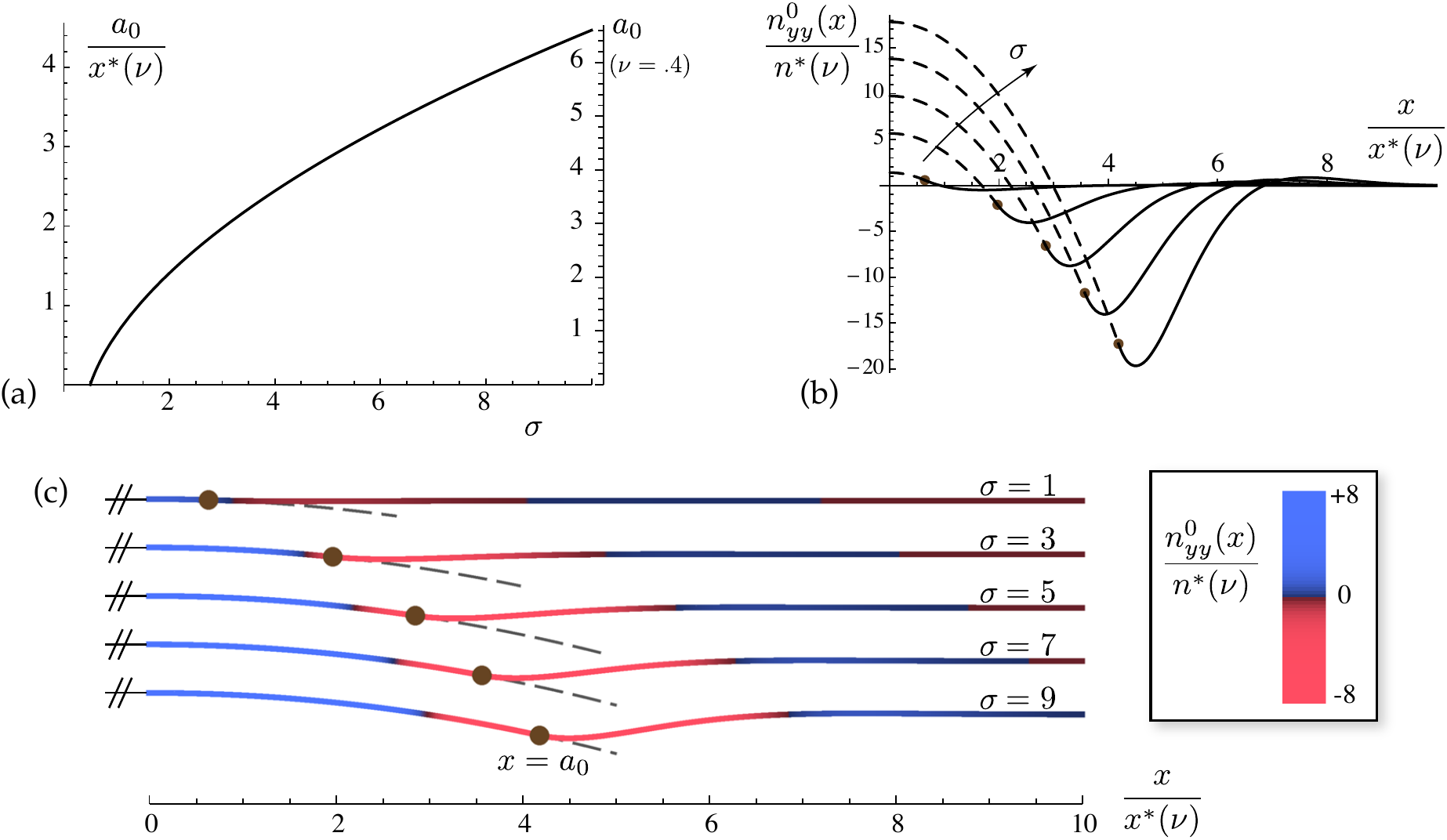}}
    \caption{Axisymmetric solution.  (a) Half-width of the band-like
    region of contact, as a function of the reduced adhesion $\sigma$.
    The vertical axis on the left-hand side shows the value of
    $a_0/x^*(\nu)$ which is independent of Poisson's ratio, and that
    on the right-hand side shows the corresponding value of $a_{0}$
    for the particular case $\nu=0.4$, as in the experiments.  (b) Hoop
    stress $n_{yy}^0(x)/n^*(\nu)$, rescaled using $n^*(\nu) =
    1/(x^*(\nu))^2$, as a function of the rescaled axial coordinate
    $x/x^*(\nu)$, for increasing values of the adhesion $\sigma$.
    When expressed in these variables, the curves are independent of 
    $\nu$.
    Dashed lines correspond to the wet region, solid lines to the free
    region.  The different curves correspond to different values of
    adhesion, namely $\sigma=1,3,5,7$ and $9$.  (c) Profile of the
    shell as obtained by cutting through a plane passing through the
    axis, for increasing values of the adhesion number $\sigma$.  Only
    one half of the cut is shown, the entire profile being symmetric
    with respect to $x=0$.  The profiles corresponding to different
    values of $\sigma$ have been offset vertically for clarity.  As
    the adhesion $\sigma$ increases, the width of the region of
    contact increases, and the hoop stress becomes more and more
    compressive around the edge of the region of contact.  This points
    to the existence of an instability, which is studied next.}
\label{fig:axi}
\end{figure}

A necessary condition for this linear system to have a solution is
\begin{equation}
\det\left(
\underline{\underline{M}}\left(\frac{a_{0}}{x^*(\nu)},\sigma\right)
\right)=0
\textrm{.}
\label{eq:det}
\end{equation}
One can simplify the determinant of $\underline{\underline{M}}$ and
rewrite this equation as
\begin{equation}
    \frac{3 + 6\,\hat{a} + 6\,\hat{a}^2 +2\,\hat{a}^3}{6\,(1 + 
    \hat{a})} = \sigma ,
    \quad\textrm{where }\hat{a} = \frac{a_{0}}{x^*(\nu)}
\label{eq:deltaa}
\end{equation}
This implicit relation, plotted in fig.~\ref{fig:axi}a, 
selects the half-width $a_{0} = (x^*(\nu)\,\hat{a}) $ of the region of
contact $\Omega^W$ in the axisymmetric configuration, as a function of
the dimensionless adhesion number $\sigma$.  Note that this half-width
$a_{0}$ goes to zero as $\sigma$ decreases to the value $1/2$: the
adhering solution disappears when the adhesion is too weak, $\sigma <
1/2$.

From now on, we assume $\sigma \geq 1/2$ and consider the value of
$a_{0}$ that is the unique solution to equation~(\ref{eq:deltaa}).
Then, the matrix in equation~(\ref{eq:sysl-Syst}) is singular and the
solutions $\underline{X}$ of the linear system span a line.
Generically, there is a unique vector $\underline{X}$ whose last
component equals the square of the quantity $a_{0}$ just found, as
imposed by equation~(\ref{eq:sysl-Unknown}).  The other components of
this particular vector $\underline{X}$ set the values of the unknowns
$A_{1}$, $A_{2}$ and $E_{0}$.  For any value of the adhesion parameter
$\sigma\geq 1/2$ this defines a unique axisymmetric solution.  Some of
these solutions are represented in figure~\ref{fig:axi}c, for
particular values of $\sigma$.  The residual stress
$\underline{\underline{n}}^0(x)$ given by
equation~(\ref{eq:prestretch}) is plotted in figures~\ref{fig:axi}b
and c.  We note that this residual hoop stress is compressive near the
edge of the region of contact, as the
film is pulled towards the axis by the adhesion.

\begin{figure}
    \centering
    \includegraphics[width=.5\textwidth]{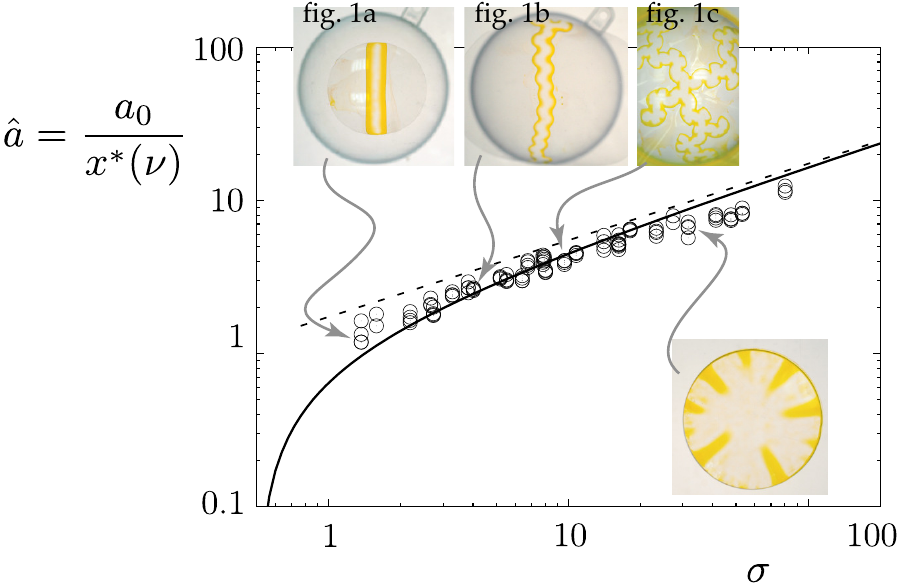}
    \caption{Size of the contact zone as a function of the adhesion
    parameter $\sigma$.  Circular symbols correspond to experimental
    data from Ref.~\protect\cite{hure2011}.  The solid line
    corresponds to the prediction of
    equation~(\protect\ref{eq:deltaa}), and the dashed
    line to the asymptotic behaviour of
    equation~(\protect\ref{eq:deltaa}).  As the solid curve comes from
    the analysis on the unbuckled stated, it cannot be expected to be
    accurate far above threshold, when $\sigma$ becomes larger than
    $\sim 10$; a good agreement is observed nevertheless.}
    \label{fig:compexp}
\end{figure}

In figure~\ref{fig:compexp}, we compare the prediction for the width
of the contact region in equation~(\ref{eq:deltaa}) to the
experimental data taken from Ref.~\cite{hure2011}, with no adjustable
parameter.
We find a good agreement.  Interestingly, the agreement is good even
in the post-buckled regime (when $\sigma$ is larger than approximately
$5$), if we measure $a_{0}$ in the experiments as half the
\emph{average} strip width; this is unexpected as the present solution
is not applicable above the bifurcation threshold.

In the limit of strong adhesion, when $\sigma$ is large, the
half-width $a_{0}$ predicted by equation~(\ref{eq:deltaa}) becomes
large as well, and is given asymptotically by $a_{0} =
\sqrt{3\,\sigma}\,x^{*}(\nu)$.  Restoring the physical units, this yields
\begin{equation}
    a_{0} \approx \sqrt{6}\, \rho \, \left(\frac{\gamma}{E h}\right)^{1/4}
    \emph{.}
\end{equation}
We recover the scaling form proposed and verified in
reference~\cite{hure2011}, and have obtained the value of the
coefficient in addition.

We have derived a family of non-linear solution of the Donnell
equations analytically that describe unbuckled, axisymmetric shapes of
the film.  The equations are non-linear and it is remarkable that
these solutions can be derived without approximation.  There is a
unique axisymmetric solution when the adhesion is large enough,
$\sigma\geq 1/2$.  For lower adhesion, $\sigma < 1/2 $ no axisymmetric
solution with extended contact can be found.

\section{Stability analysis}
\label{sectstab}

Given the presence of compressive hoop stress in the axisymmetric
solution shown in figure~\ref{fig:axi}c, the neighborhood of the edges
of the region of contact can become unstable, especially when the
adhesion number $\sigma$ becomes large.  Making use of the explicit
solution for the unbuckled state, one can approach this question by
studying the linear stability of the unbuckled state.  This is the
goal of the present section.

\subsection{Perturbations}

We investigate the presence of bifurcated branches near the unbuckled
state, and introduce a perturbation of the previous solution in the
form:
\begin{subequations}
\label{eq:perturb}
\begin{alignat}{2}
    u_{x}(x,y) &= u_{x}^{0i}(x) & + & u_{x}^{1i}(x)\,\cos{(k\,y)}
    \label{eq:perturb-ux}
    \\
    u_{y}(x,y) &= E_{0} \, y & + & u_{y}^{1i}(x)\,\sin{(k\,y)}
    \label{eq:perturb-uy}
    \\
    w(x,y) &= w_0^i(x) & + & w_1^i(x)\,\cos{(k\,y)}
    \textrm{.}
    \label{eq:perturb-w}
\end{alignat}
\end{subequations}
Here the index $1$ refers to perturbations from the axisymmetric
state.  As the axisymmetric solution is invariant in the $y$
direction, the harmonic dependence of the perturbations on the $y$
variable introduced above is the only one that we need to consider: a
generic perturbation can be recovered by linear superposition.

The edge $G$ of the wet region may deform upon the instability.
Therefore we introduce a perturbation of its boundary,
\begin{equation}
        g_{+}(y) = a_{0} + a_{1}\, \cos(k\,y)
	\textrm{.}
	\label{eq:G+Perturbation}
\end{equation}
Since the base solution is mirror-symmetric with respect to the plane
$x=0$, we shall only need to consider perturbations that are either
symmetric, or antisymmetric.  The benefit is that we only need to
solve the linearized equations on half the domain, $x\geq 0$, using
initial conditions at the center of symmetry $x=0$ that reflect the
type of symmetry under consideration --- details will be provided
below.  The shape of the boundary at $x=-a_{0}$ can be reconstructed
from the other boundary using equation~(\ref{eq:G+Perturbation}):
$g_{-}(y) = - a_{0} - a_{1}\, \cos(k\,y)$ in the symmetric case (also
called the varicose pattern, as the width of the wet region gets
modulated while its center-line remains straight), and by $g_{+}(y) =
-a_{0} + a_{1}\, \cos(k\,y)$ in the antisymmetric case (in this
sinuous mode, the width of the wet region remains constant to first
order but its center-line undulates laterally), see
figure~\ref{fig:sympert}.

\subsection{Linearized equilibrium in the interiors of the domains}

Inserting the expansions~(\ref{eq:perturb-ux}--\ref{eq:perturb-w})
into the equations of equilibrium~(\ref{eq:equilibrium}), and
linearizing around the unbuckled state $(u_{x}^{0i}, E_{0}, w_0^i,
a_0)$, yields three coupled, linear ordinary differential equations
for the functions $u_x^{1i}(x)$, $u_y^{1i}(x)$, $w_1^i(x)$.  These
equations are fourth order with respect to $w_1^i$ and second order
with respect to $u_x^{1i}$ and $u_y^{1i}$.  

In the wet region $\Omega_W$, the deflection is prescribed by
equation~(\ref{eq:sphereParabolicApproximDimensionalForm}) and
$w_1^W(x)=0$.  We are not interested in computing the perturbation to
the contact pressure, and will not use the transverse
equilibrium~(\ref{eq:shellEqTransverse}) there: we are left with the
linearized equations for in-plane equilibrium, which are two second
order, ordinary differential equations for $u_x^{1i}$ and $u_y^{1i}$.

The linearized
equations of equilibrium can be cast into an equivalent first-order 
form by introducing the state vectors, defined in each region by
\begin{subequations}
    \begin{align}
	\underline{U}_W(x) & = 
	[u_x^{1W}, u_x^{1W'}, u_y^{1W}, u_y^{1W'}] \\
	\underline{U}_F(x) & = 
	[u_x^{1F}, u_x^{1F'}, u_y^{1F}, u_y^{1F'}, 
	w_1^F, w_1^{F'}, w_1^{F''}, w_1^{F'''}]
	\textrm{.}
    \end{align}
\end{subequations}
The linearized equilibrium in the interior of each domain then reads
\begin{subequations}
    \label{eq:systperturb}
    \begin{alignat}{2}
	\underline{U}_W'(x) & = \underline{\underline{A}}_W(\nu, 
	\sigma,k,x)\cdot\underline{U}_W(x) \ \ \ \ \ &\mathrm{for }&\ \ x \in \Omega_W
	\label{eq:systperturb1} \\
	\underline{U}_F'(x) & = \underline{\underline{A}}_F(\nu,\sigma,k,x)\cdot\underline{U}_F(x) \ \ \ \ \ &\mathrm{for }&\ \  x \in \Omega_F
	\textrm{,}
	\label{eq:systperturb2}
    \end{alignat}
\end{subequations} 
where $\underline{\underline{A}}_W$ (respectively
$\underline{\underline{A}}_F$) is a $4\times 4$ (respectively $8\times
8$) matrix whose coefficients depend on the dimensionless coordinate
$x$, on the adhesion parameter $\sigma$, on Poisson's ratio $\nu$,
as well as on the wave number $k$.  This dependence arises either
directly, or indirectly through the unbuckled solution
$(u_x^{0i},E_{0}, w_0^i, a_0)$.

\subsection{Asymptotic behavior far from the sphere}

We first integrate the linearized equations of equilibrium along the
free region, $a_{0} \leq x \leq L_{x}$.  We start from the free end
$x=L_{x}$ where we use initial conditions consistent with the
stressfree boundary conditions, and proceed towards the moving
interface $x=a_{0}$.  The perturbation computed near $x=a_{0}$ will be
ultimately reconciled with that coming from the wet region using the
equations at the mobile interface.

We consider an infinitely long shell, $L_x \rightarrow +\infty$.  This
is an accurate approximation as in the experiments the film is much
wider than the region of contact.  We must only consider
solutions of the linearized problem that remain bounded for large $x$:
we start by studying their asymptotic behavior, which generically is
exponential.  The trial form $u_x^{1F}(x) = U_x^{1F} \exp{(\tau x)}$,
$u_y^{1F}(x) = U_y^{1F} \exp{(\tau x)}$ and $w_1^F(x) = W_1^F
\exp{(\tau x)}$ is inserted into the linearized
equilibrium~(\ref{eq:systperturb2}).  Denoting
$\underline{U}_{\infty}^F = [U_x^{1F}, U_y^{1F}, W_1^F]$ the vector
collecting the unknown amplitudes, the equations of equilibrium are 
asymptotically satisfied provided the following condition holds:
\begin{equation}
    \underline{\underline{A}}_{\infty} (\nu, \sigma,k,\tau)\cdot \underline{U}_{\infty}^F = \underline{0}
    \textrm{,}
\label{ainfini}
\end{equation}
where $\underline{\underline{A}}_{\infty}$ captures the asymptotic
form of the linearized equations, 
\begin{equation}
\underline{\underline{A}}_{\infty} (\nu, \sigma,k,\tau) = \begin{pmatrix}
\frac{k^2}{2}(-1 + \nu) + \tau^2  & \frac{k}{2}(1+\nu)\tau & \nu \tau\\
 - \frac{k}{2}(1+\nu) \tau & -k^2 - \frac{1}{2}(-1+\nu)\tau^2 & -k\\
 -\nu\tau & -k & -1 -k^4 -1 -k^4 + 2k^2\tau^2 - \tau^4 
\end{pmatrix}
\textrm{.}
\end{equation}
The acceptable values of the decay rate $\tau$ are found by requiring 
that equation~(\ref{ainfini}) has non-trivial solutions:
\begin{equation}
    \det{\underline{\underline{A}}_{\infty}} 
    = \frac{1}{2}(-1+\nu)(k^8 - 4k^6\tau^2 + \tau^4 + 6k^4 \tau^4 - \nu^2 \tau^4 - 4 k^2 \tau^6 + \tau^8)     
    = 0
    \textrm{.}
\end{equation}
Each one of the eight complex roots, denoted $\tau_i$, is associated
with an eigenvector $\underline{U}_{\infty}^{Fi}$.  

The stressfree boundary conditions~(\ref{eq:boud}) will be
automatically satisfied, provided the perturbation stays bounded for
large $x$.  Therefore, out of the eight possible exponential
behaviors, we can keep all four non-divergent solutions which are such
that the real part of $\tau_j$ is negative, $\Re(\tau_j) < 0$.  By
convention, these values of $\tau_j$ are indexed by $1\leq j\leq 4$.
A generic, non-divergent solution of the linearized
equilibrium~(\ref{eq:systperturb2}) is then found by linear
superposition,
\begin{equation}
    \label{eq:asymptsol}
    \underline{U}_F(x_{m}) \approx \sum_{j=1}^4 i_j\,
    \underline{U}_{\infty}^{Fj}\,e^{\tau_j\, x_m}
    \textrm{,}
\end{equation}
an approximation that is accurate for large values of $x_{m}$.  The
four unknown amplitudes of the converging modes are collected into a
vector $\underline{I}_F = [i_1,i_2,i_3,i_4]$.

We can now use the asymptotic form~(\ref{eq:asymptsol}) as an initial
value, and integrate the linearized equation towards the edge of the
free region, $x = a_{0}$.  By linearity, the state vector
$\underline{U}_F(a_0^+)$ depends linearly on the asymptotic amplitudes
$\underline{I}_F$:
\begin{equation}
    \underline{U}_F(a_0^+) = 
    \underline{\underline{\mathcal{S}}}_F(\nu,\sigma,k)\cdot \underline{I}_F
    \textrm{.}
\label{SF}
\end{equation}
This $\underline{\underline{\mathcal{S}}}_F$ is the so-called shoot
matrix $\underline{\underline{\mathcal{S}}}_F$, and its size is
$8\times 4$.  Its columns are computed as follows.  First the initial
condition for the linearized equilibrium~(\ref{eq:systperturb2}) are
computed based on the asymptotic form of the
perturbation~(\ref{eq:asymptsol}), using a finite but numerically
large value of $x_{m}$; all the coefficients $i_{j}$'s are set to
zero except for one which is set to one.  The linearized equilibrium
is then integrated from $x=x_{m}$ to $x=a_{0}$ numerically, and the
final state vector $\underline{U}_F(a_0)$ is used to fill in the
corresponding column of $\underline{\underline{\mathcal{S}}}_F$.  This
shoot matrix captures the linearized response of the free region,
including the remote stressfree edge, as viewed from the edge $G$.

\subsection{Symmetry conditions at the center of the domain}

\begin{figure}
\begin{center}
    \begin{minipage}[c]{0.49\textwidth}
	\subfigure[]{\includegraphics[width=7cm]{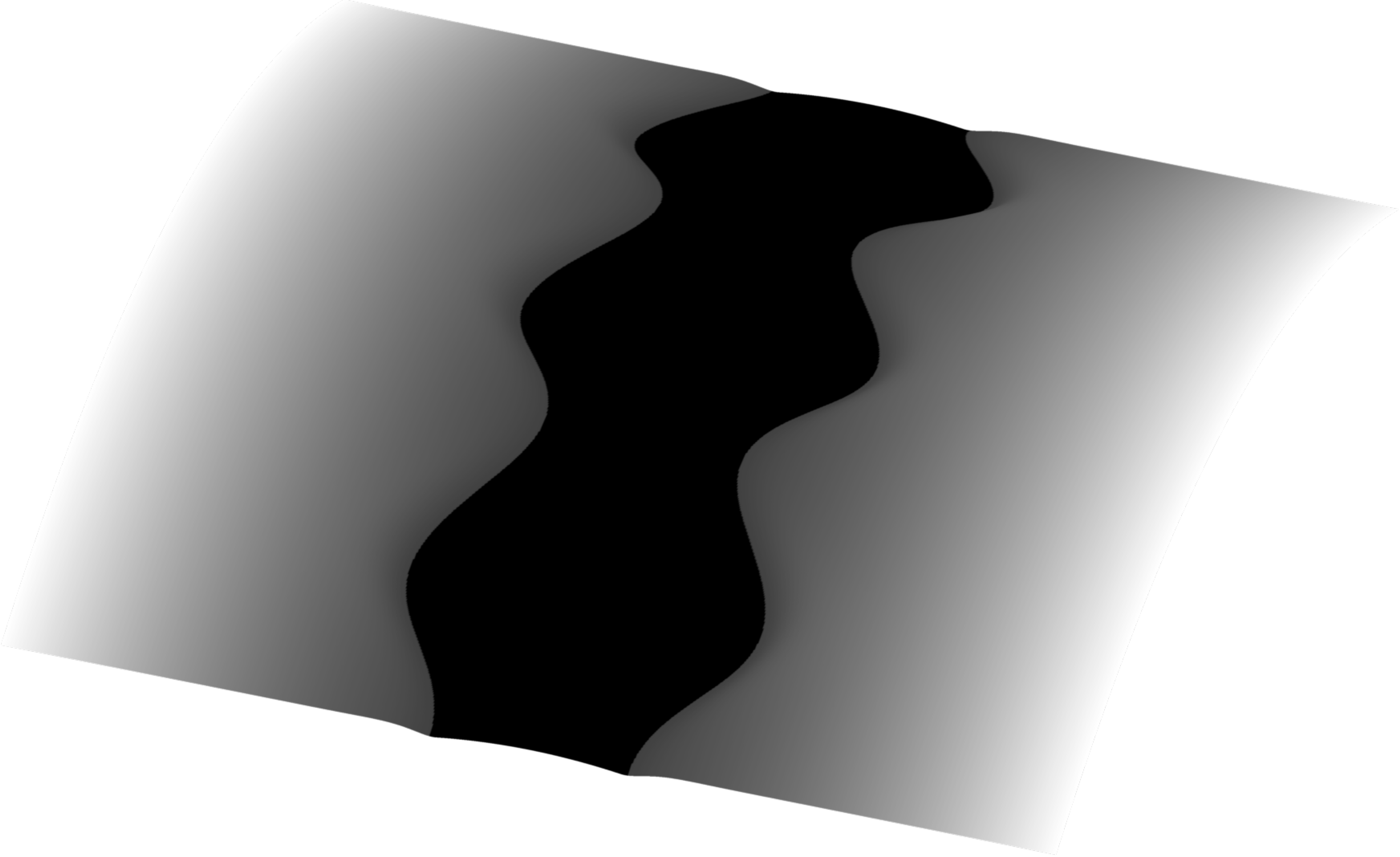}}
    \end{minipage}
    \begin{minipage}[c]{0.49\textwidth}
	\subfigure[]{\includegraphics[width=7cm]{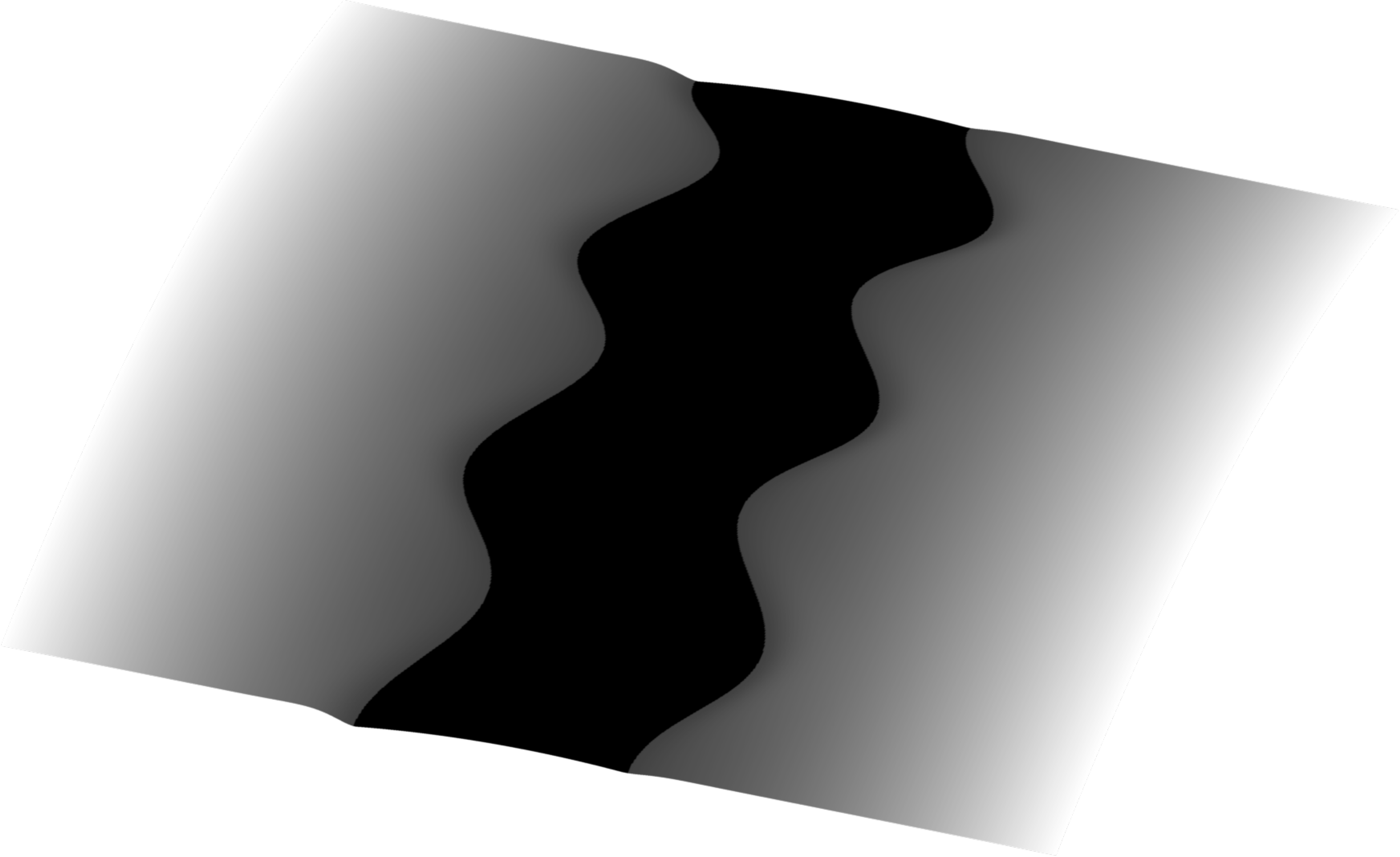}}
    \end{minipage}
\end{center}
\caption{Two types of buckling modes: (a) Symmetric (varicose) and (b)
antisymmetric (sinuous) perturbations.}
    \label{fig:sympert}
\end{figure}

As explained earlier, the symmetry of the base solution by a mirror
reflection changing $x$ to $(-x)$ allows us to consider perturbations
that are either symmetric or antisymmetric, while retaining full
generality.  These two types of buckling modes are depicted in
figure~\ref{fig:sympert}.

The type of symmetry dictates the initial condition at the center of
symmetry, $x=0$.  A varicose perturbation, shown in figure
\ref{fig:sympert}a, is denoted by a superscript $(+)$ as it is symmetric
with respect to the $x$ axis; for this type of symmetry, the
conditions $u_x^{1W}(0) = u_y^{1W'}(0) = 0$ hold.  A sinuous
perturbation, shown in figure \ref{fig:sympert}b, is denoted by a
superscript $(-)$ as it is antisymmetric with respect to the $x$ axis;
for this other type of symmetry, the conditions $u_x^{1W'}(0) =
u_y^{1W}(0) = 0$ hold.  We can thus define two independent initial
state vectors in the wet region as follows: for the analysis of the
varicose mode, $\underline{U}_W^{+,1}(0) = [0,1,0,0]$ and
$\underline{U}_W^{+,2}(0) = [0,0,1,0]$.  For the analysis of the
sinuous mode, $\underline{U}_W^{-,1}(0) = [1,0,0,0]$ and
$\underline{U}_W^{-,2}(0) = [0,0,0,1]$.  A generic initial condition
compatible with the symmetry is obtained by linear superposition,
using two unknown amplitudes which we denote $i_{5}$ and $i_{6}$:
\begin{equation} 
    \label{eq:0sol}
    \underline{U}_W^{\pm}(0) = \sum_{j=1}^2 i_{j+4}\,\underline{U}_W^{\pm,j}(0)
    \textrm{.}
\end{equation}
Integrating equation~(\ref{eq:systperturb1}) across the wet region
$\Omega_F$ using each of the various modes in equation~(\ref{eq:0sol})
successively, we compute a shoot matrix for the wet region:
\begin{equation}
    \underline{U}_W^{\pm}(a_0^-) = 
    \underline{\underline{\mathcal{S}}}_W^{\pm}(\nu,\sigma,k)\cdot \underline{I}_W
    \textrm{.} 
\label{SW}
\end{equation}
There are in fact two such shoot matrices, one for each type of
symmetry, and with size $4\times 2$ each.

\subsection{Assembly}

Let us define an assembled state vector at the boundary $x=a_{0}$ by collecting those
relevant to the free and wet regions:
\begin{equation}
    \underline{U}(a_0) = [\underline{U}_F(a_0^+), \underline{U}_W^{\pm}(a_0^-),a_1]
    \textrm{.}
    \label{eq:assembledStateVector}
\end{equation}
We have also appended the perturbation $a_{1}$ of the boundary shape,
which will soon be needed.

We can similarly define an assembled shoot vector by
\begin{equation}
    \underline{I} = [\underline{I}_F,\underline{I}_W,a_1]
    \textrm{.}
    \label{eq:assembledShootingVector}
\end{equation}
This $\underline{I}$ is the main unknown of our stability problem, and
will be shown to satisfy an eigenvalue problem.

Then equations~(\ref{SF}) and (\ref{SW}) can be rewritten in compact
form as:
\begin{equation}
    \underline{U}(a_0) = \underline{\underline{\mathcal{S}}}_{WF}^{\pm}(\nu,\sigma,k) \cdot \underline{I}
\label{WFA0}
\end{equation}
where the global shoot matrix
$\underline{\underline{\mathcal{S}}}_{WF}^{\pm}$ is formed by 
assembling into blocks the shoot matrices previously computed:
\begin{equation}
    \underline{\underline{\mathcal{S}}}_{WF}^{\pm}(\nu,\sigma,k) = \begin{pmatrix}
    \underline{\underline{\mathcal{S}}}_{F} & 0 & 0 \\
    0 & \underline{\underline{\mathcal{S}}}_{W}^{\pm} & 0 \\
    0 & 0 & 1
\end{pmatrix}
\textrm{.}
\end{equation}
Its size is $13\times 7$.

\subsection{Equilibrium of the moving boundary}

At this point, we are ready to close the formulation of the stability
problem by using the remaining continuity and jump conditions at the
edge of the contact region.  Upon linearization, the seven
conditions~(\ref{eq:eqcontinuity}) can be written in matrix notation
as:
\begin{equation}
    \underline{\underline{\mathcal{C}}}(\nu,\sigma, k)\cdot \underline{U}(a_0) = \underline{0}
    \emph{.}
\label{continua0}
\end{equation}
The matrix $\underline{\underline{\mathcal{C}}}$ is of size $7\times
7$.  It collects the coefficients appearing in these linearized
equations.  The equations~(\ref{eq:eqcontinuity}) hold on the mobile
curve $x=g_{+}(x)$ which is perturbed according to
equation~(\ref{eq:G+Perturbation}).  As a result, the linearized
equations have terms proportional to the boundary perturbation $a_{1}$
times the gradients of the base solution, and these terms are used to
fill the last column of $\underline{\underline{\mathcal{C}}}$.  This
allows the perturbation to the boundary to be treated without
approximation.

Note that the inhomogeneous term $(-2\,\sigma)$ appearing in the
right-hand side of the adhesion condition~(\ref{eq:mNNJump})
disappears upon linearization.

\subsection{Linear stability formulated as an eigenproblem}

By combining the equilibrium of the interface in 
equation~(\ref{continua0}) and the integration in the free and wet 
domain captured in equation~(\ref{WFA0}), we can write the 
linear stability problem as an eigenvalue problem,
\begin{equation}
    \big[ 
    \underline{\underline{\mathcal{C}}} (\nu,\sigma, k)
    \cdot
    \underline{\underline{\mathcal{S}}}_{WF}^{\pm}(\nu,\sigma,k) 
    \big]
    \cdot
    \underline{I} = \underline{0}
    \textrm{.}
\end{equation}
 
The existence of a linearly unstable mode requires
\begin{equation}
    \det{\big[ \underline{\underline{\mathcal{C}}}(\nu,\sigma, k)
    \cdot 
    \underline{\underline{\mathcal{S}}}_{WF}^{\pm}(\nu,\sigma,k) \big]} = 0
    \textrm{.}
    \label{eq:linearStabilityInSymbolicForm}
\end{equation}
This is an implicit equation for the wave number $k$ as a function of
the adhesion parameter $\sigma$ and of Poisson's ratio $\nu$.  The
corresponding eigenvector $\underline{I}$ allows the linearly unstable
mode to be reconstructed.

\subsection{Numerical results}

\begin{figure}
    \centerline{\includegraphics[width=.99\textwidth]{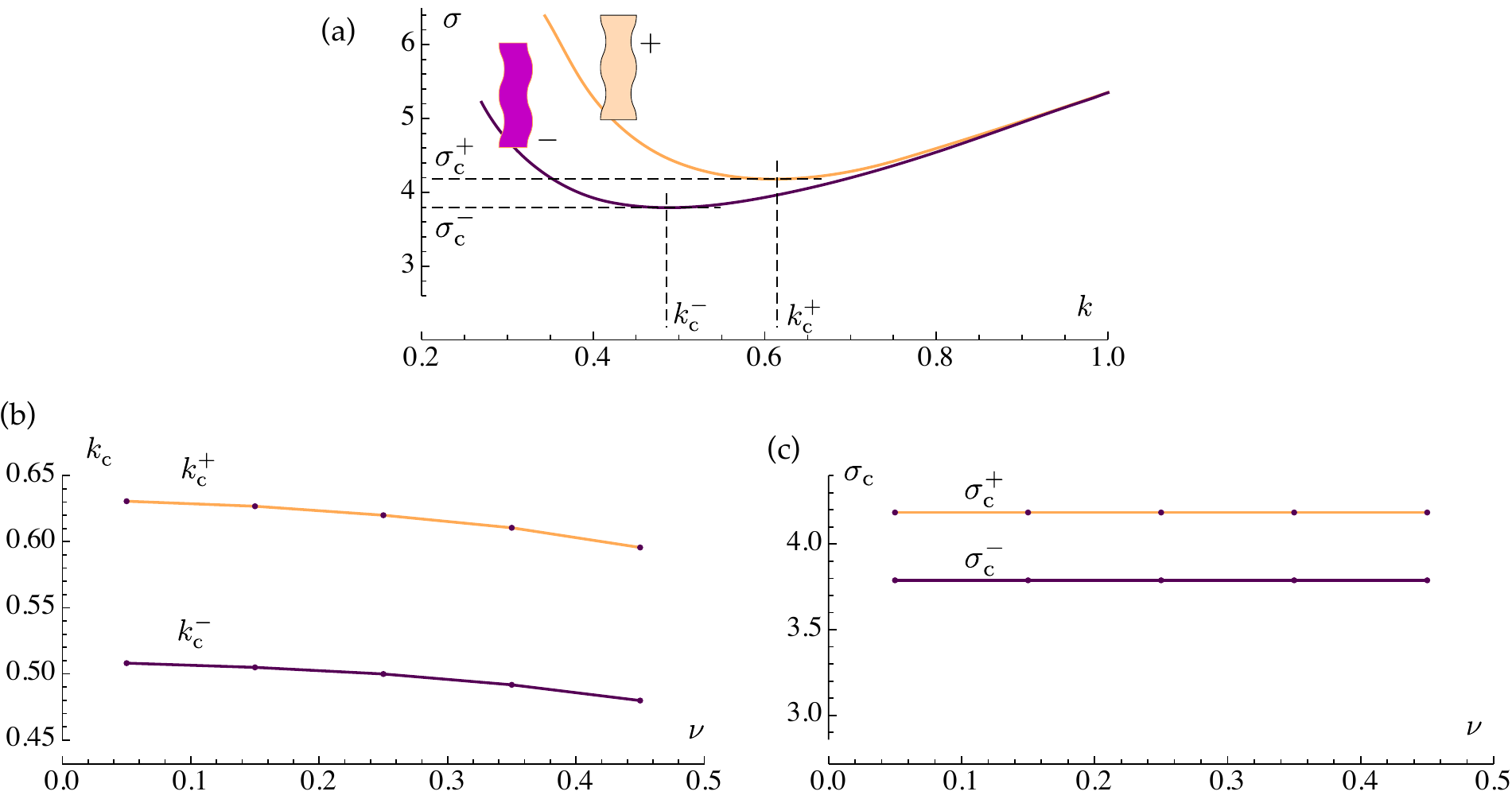}}
    \caption{(a) Linear stability diagram for $\nu=0.4$.  The most
    unstable symmetric and antisymmetric modes are defined by
    $\sigma_c^{\pm}$ and $k_c^{\pm}$ (b,c) Dependence of the critical
    wavevector $k_c^{\pm}$ and adhesion number $\sigma_c^{\pm}$ on
    Poisson's ratio.}
    \protect\label{fig:linStabilityResults}
\end{figure}

We explained how the shoot matrix
$\underline{\underline{\mathcal{S}}}_{WF}^{\pm}(\nu,\sigma,k)$ and the
linearized equilibrium of the boundary
$\underline{\underline{\mathcal{C}}}(\nu,\sigma, k)$ can be computed
numerically for specific values of their arguments.  For a given value
of Poisson's ratio $\nu$ and for each type of symmetry, we repeatedly
computed the determinant for different values of $\sigma$ and $k$, and
then plotted the implicit curve defined by
equation~(\ref{eq:linearStabilityInSymbolicForm}) in the plane
$(k,\sigma)$.  The result is shown in
figure~\ref{fig:linStabilityResults}a for a typical value of Poisson's
ratio, $\nu=0.4$.
Numerical convergence with respect to the parameter $x_{m}$ required
in the calculation of the shoot matrix
$\underline{\underline{\mathcal{S}}}_F$ was attested by the fact that
further increase of $x_{m}$ did not significantly affect the
stability curves; typically, this required $x_{m}\approx 12$.

The most unstable wavenumber $k_{\mathrm{c}}^\pm$ and the critical
adhesion $\sigma_{\mathrm{c}}^\pm$ correspond to the minimum of each
of the curves in the plane $(k,\sigma)$.  For each type of symmetry,
the most unstable mode depends only on Poisson's ratio, hence the
notations $\sigma_{\mathrm{c}}^\pm(\nu)$ and
$k_{\mathrm{c}}^\pm(\nu)$.  The corresponding curves are plotted in
figure~\ref{fig:linStabilityResults}b and c.

The stability analysis predicts that the most
unstable mode is always the antisymmetric (sinuous) one.  Indeed, the
curve corresponding to this mode is always below the other curve in
figure~\ref{fig:linStabilityResults}c.  This is consistent with the
experiments, where the varicose mode has never been observed.

It turns out that the dependence of both
$\sigma_{\mathrm{c}^\pm}(\nu)$ and $k_{\mathrm{c}}^\pm(\nu)$ on
Poisson's ratio can be captured by simple formulas which match the
numerical results perfectly, within numerical accuracy:
\begin{subequations}
    \label{eq:almostAnalyticalResultsForStability}
\begin{alignat}{2}
     k_{\mathrm{c}}^-(\nu)&\approx  \frac{0.718}{x^*(\nu)}
     &\qquad
     \sigma_{\mathrm{c}}^-(\nu)& \approx  3.788  
     \textrm{,}
     \\
    k_{\mathrm{c}}^+(\nu)&\approx \frac{0.891}{x^*(\nu)}
    &\qquad
    \sigma_{\mathrm{c}}^+(\nu)&\approx  4.184
    \textrm{,}
\end{alignat}
\end{subequations}
where $x^*(\nu)$ is the function defined in
equation~(\ref{eq:defXStar}).  We have no explanation to offer for the
fact that $\sigma_{\mathrm{c}}^\pm$ is independent of $\nu$, and that
$k_{\mathrm{c}}^\pm(\nu)$ depends on $\nu$ as $1/x^*(\nu)$.  This
probably points to the fact that a proper rescaling of the various
quantities would allow the parameter $\nu$ to be removed from the
linearized equations altogether.  We have not been able, however, to
identify such a rescaling.

The predictions relevant to the experiments, where we used a film 
with Poisson's ratio $\nu= 0.4$, are
\begin{equation}
    \sigma_{\mathrm{c}}^-(0.4) = 3.788,
    \qquad
    k_{\mathrm{c}}^-(0.4) = 0.486,
    \qquad
    \ell_{\mathrm{c}}^- = \frac{2\,\pi}{k_{\mathrm{c}^-}(0.4)} = 12.93
    \textrm{.}
    \label{eq:PredictionsForNu=.4}
\end{equation}
Here $\ell_{\mathrm{c}}^-$ denotes the wavelength of the sinuous
instability.  These results are in qualitative agreement with the
picture shown in figure~\ref{fig:ExpPRL}: the straight contact region
looses stability somewhere between $\sigma=1.37$ and $\sigma=4.01$,
and the buckled pattern is indeed sinuous (antisymmetric).  A detailed
and systematic comparison to the experiments is presented in the
following section, using a novel experimental setup that
allows the adhesion number $\sigma$ to be varied continuously.
     
\section{Comparison to experiments}
\label{sectcompexp}

\begin{figure}
    \begin{center}
	\includegraphics[height=9cm]{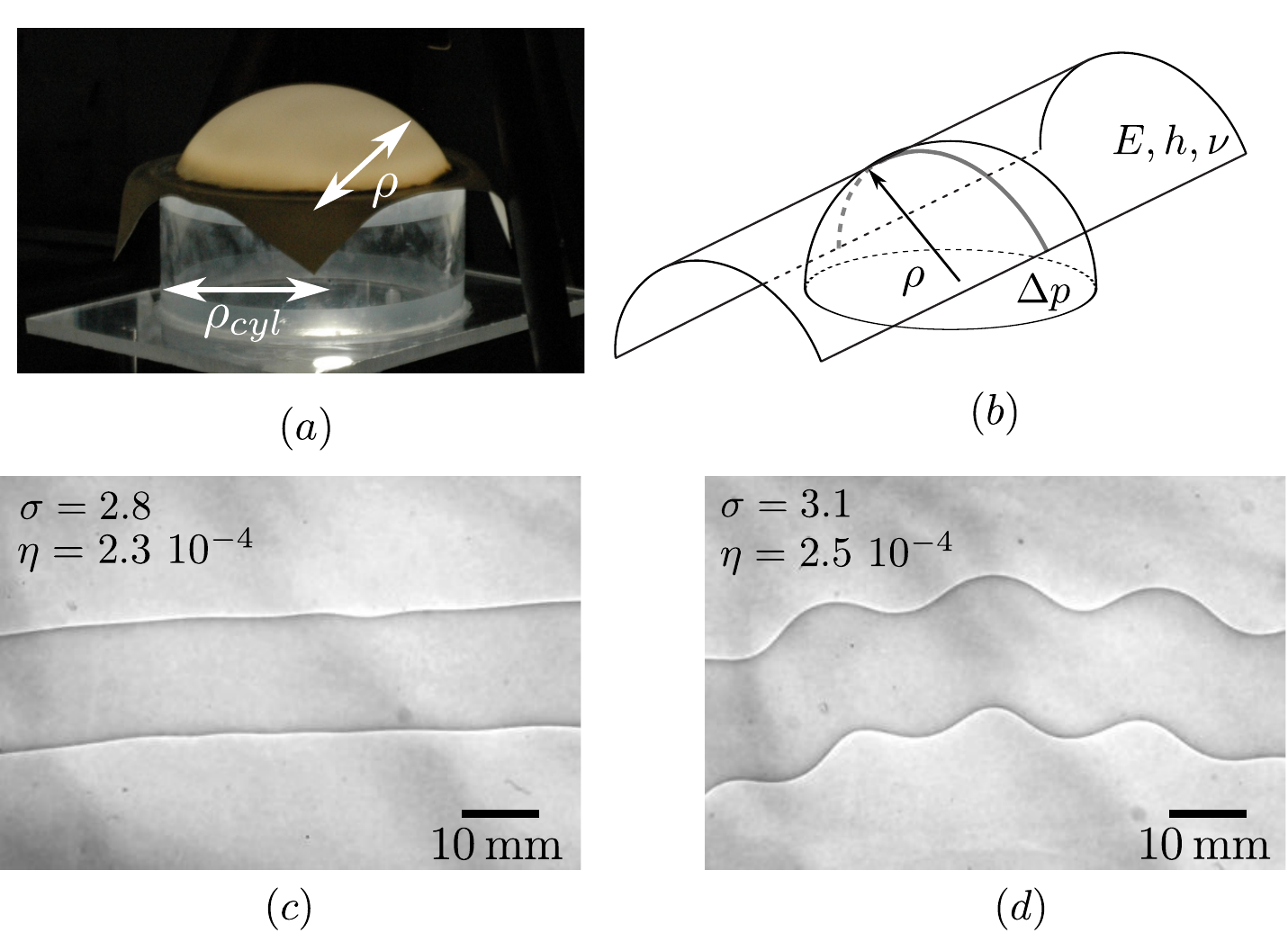}
    \end{center}
    \caption{(a,b) Experimental setup.  A square latex sheet glued on
    top of a rigid cylinder is inflated by a pressure $\Delta p$ and
    deforms into a spherical cap.  It is coated with ethanol and a
    thin film is applied onto it.  (c,d) Experimental images of the
    adhering regions, as viewed from top: the darker strip
    corresponds to the contact zone.  Decreasing the pressure $\Delta
    p$ increases the adhesion parameter $\sigma$, and induces a
    transition from straight to oscillatory edges.}
    \label{fig:setup}
\end{figure}

In order to check the predictions of the linear stability analysis, we
have developed a novel experimental setup that allows the adhesion
parameter $\sigma$ to be continuously varied while the shape of the
contact area is monitored.  To do so, we replaced the rigid spheres
used in previous work~\cite{hure2011} by an inflatable latex membrane.
The membrane was cut out in a latex sheet with Young's modulus
$E_{\mathrm{m}} = 1.2~\mathrm{MPa}$ and thickness
$h_{\mathrm{m}}=0.6~\mathrm{mm}$, and glued on top of vertical rigid
cylinders with radius $\rho_{\mathrm{cyl}} =
20,38,100,295~\mathrm{mm}$, see figure~\ref{fig:setup}a.
Increasing the pressure by $\Delta p$ inside the cylindrical container
allows the radius of the latex sheet to be varied continuously from
$\rho = \infty$ when the membrane is flat, to $\rho =
\rho_{\mathrm{cyl}}$ when it is half a sphere; the adhesion parameter
$\sigma$ then varies according to equation~(\ref{eq:delta2}).  We
checked that the shape of the membrane is almost spherical.  The value
of the radius of curvature $\rho(\Delta p)$ was measured from
pictures taken from the side.

The spherical cap was then coated with ethanol, whose surface tension
is $\gamma=22.4~\mathrm{mN}\,\mathrm{m}^{-1}$.  Thin polypropylene
films were applied on top of it, as shown in figure~\ref{fig:setup}b.
The films, produced by Innovia films, have a Young's modulus $E=2.6
\pm 0.2~\mathrm{GPa}$, a Poisson's ratio $\nu=0.4$, and their
thickness ranges from $h=15\,\mu\mathrm{m}$ to $90\,\mu\mathrm{m}$.
Ethanol was chosen because it wets both latex and polypropylene, and
because its surface tension is not very sensitive to impurities.
Snapshots are simultaneously taken from top and from the side to
monitor the shape of the pattern as a function of the radius of the
spherical cap.  The transition from band-like contact pattern to
sinuous pattern was observed as a result of membrane deflation,
see figure~\ref{fig:setup}c,d.

\begin{figure}
    \begin{minipage}[b]{0.5\textwidth}
	\subfigure[]{\includegraphics[height=5cm]{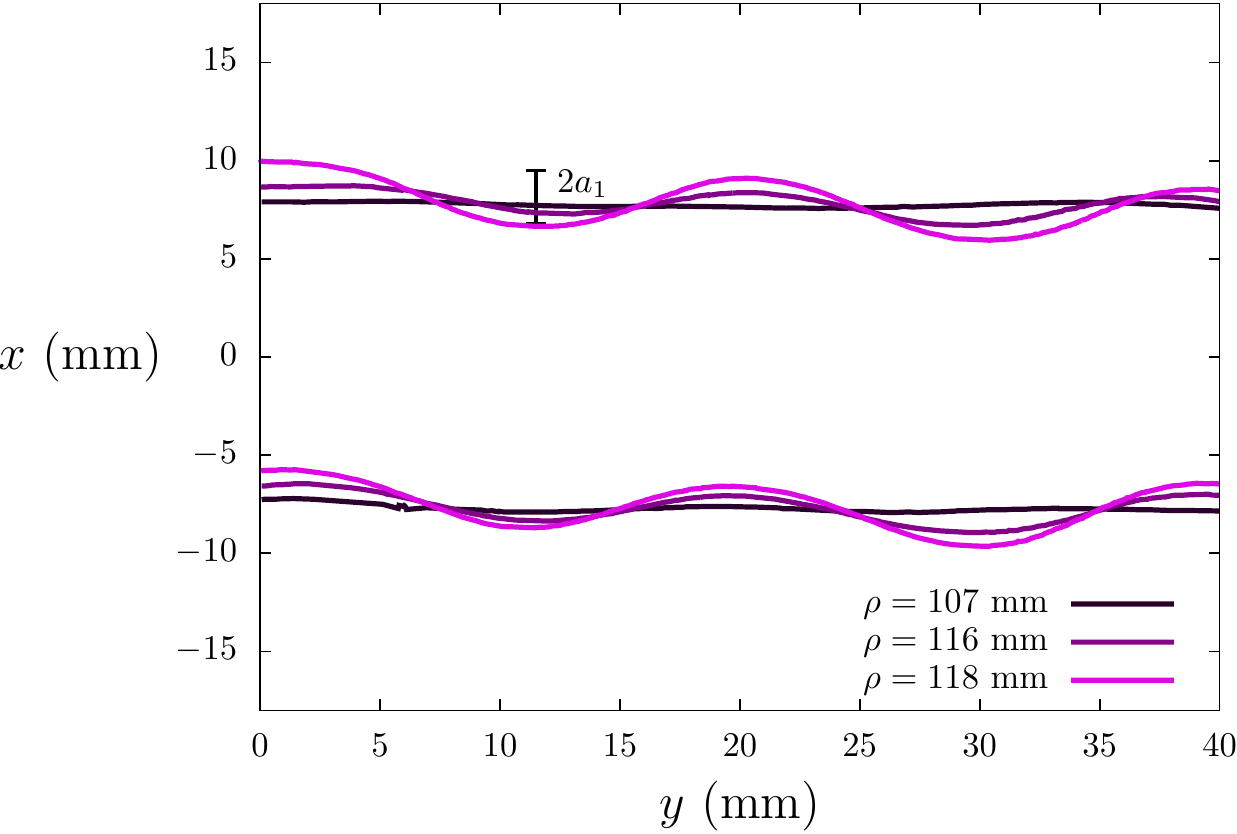}}
    \end{minipage}
    \begin{minipage}[b]{0.5\textwidth}
	\subfigure[]{\includegraphics[height=5cm]{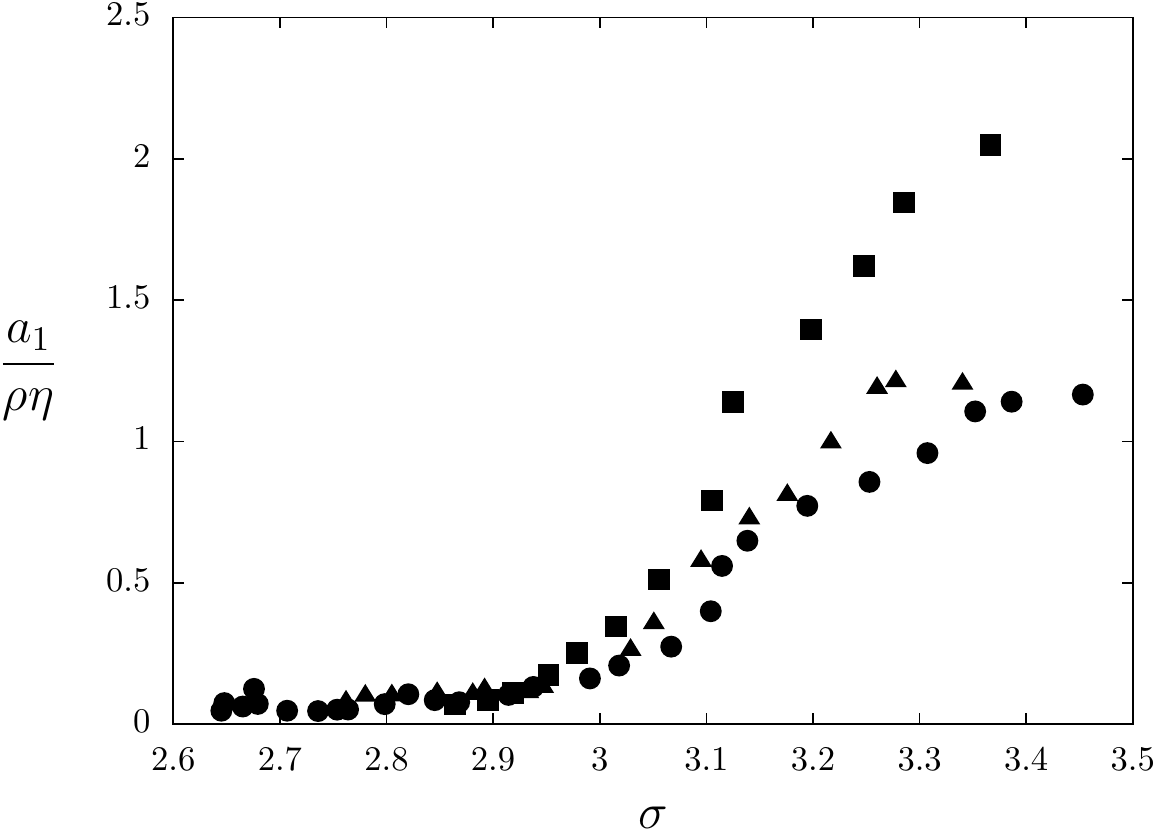}}
    \end{minipage}
    \caption{(a) Evolution of the edges as a function of the radius of
    the spherical cap $\rho$.  All curves correspond to the same film,
    $h=50~\mu\mathrm{m}$.  (b,c) Evolution of the rescaled amplitude
    $a_{1}$ of the edge profile as a function of the reduced adhesion
    number $\sigma$.  Variations of $\sigma$ have been achieved by
    deflating the membrane.  The experiment has been repeated three
    times using the same film with thickness $h=50~\mu\mathrm{m}$, and
    the symbols correspond to the different experiments.  The value of
    the threshold is very reproducible, but the amplitude of buckling
    in the post-buckled regime is more scattered.}
\label{fig:instability}
\end{figure}

The typical evolution of the edges profile as a function of the radius
$\rho$ is shown in figure~\ref{fig:instability}a. 
The edges remain straight until the sphere reaches a critical radius
$\rho_c$ and then become undulatory; the amplitude of undulation
increases as $\rho$ is further increased.  The observed buckled
patterns are always sinuous, as predicted by the stability analysis.
By fitting the edges profile with a cosine function $a_1
\cos{(kx+\phi)}$, we extracted the wavelength and the amplitude of the
oscillations.  Our measurements for a particular polypropylene film
($h=50~\mu\mathrm{m}$) are collected in figure~\ref{fig:instability}b,
which shows the dependence of the amplitude of oscillation on the
sphere radius.  The amplitude of the undulations goes to zero at the
instability threshold, which is consistent with a supercritical
bifurcation.  When the experiment is repeated several times using the
same film, the value of the bifurcation threshold is found to be quite
reproducible.  The scattering of the data for the buckling amplitude
well above the threshold may attributed to variations in the volume
of ethanol used, to friction between the film and the sphere, and to
variations in the way that the film is initially laid out on the
sphere.

Restoring the physical units, we rewrite the predictions of the
analysis of linear stability in
equation~(\ref{eq:PredictionsForNu=.4}) as
\begin{equation}
    \rho_{\mathrm{c}}(0.4) = 3.788\, L_{ec},
    \qquad
    \ell_{\mathrm{c}}(0.4) = 13.52\, \sqrt{h \, L_{ec}}
    \textrm{.}
    \label{eq:PredictionsForNu=.4exp}
\end{equation}
We recall that $\rho_{\mathrm{c}}$ and $\ell_{\mathrm{c}}$ denote the
radius of curvature of the sphere, and the wavelength of the pattern at
threshold, respectively.

\begin{figure}
    \begin{minipage}[b]{0.49\textwidth}
	\subfigure[]{\includegraphics[height=5cm]{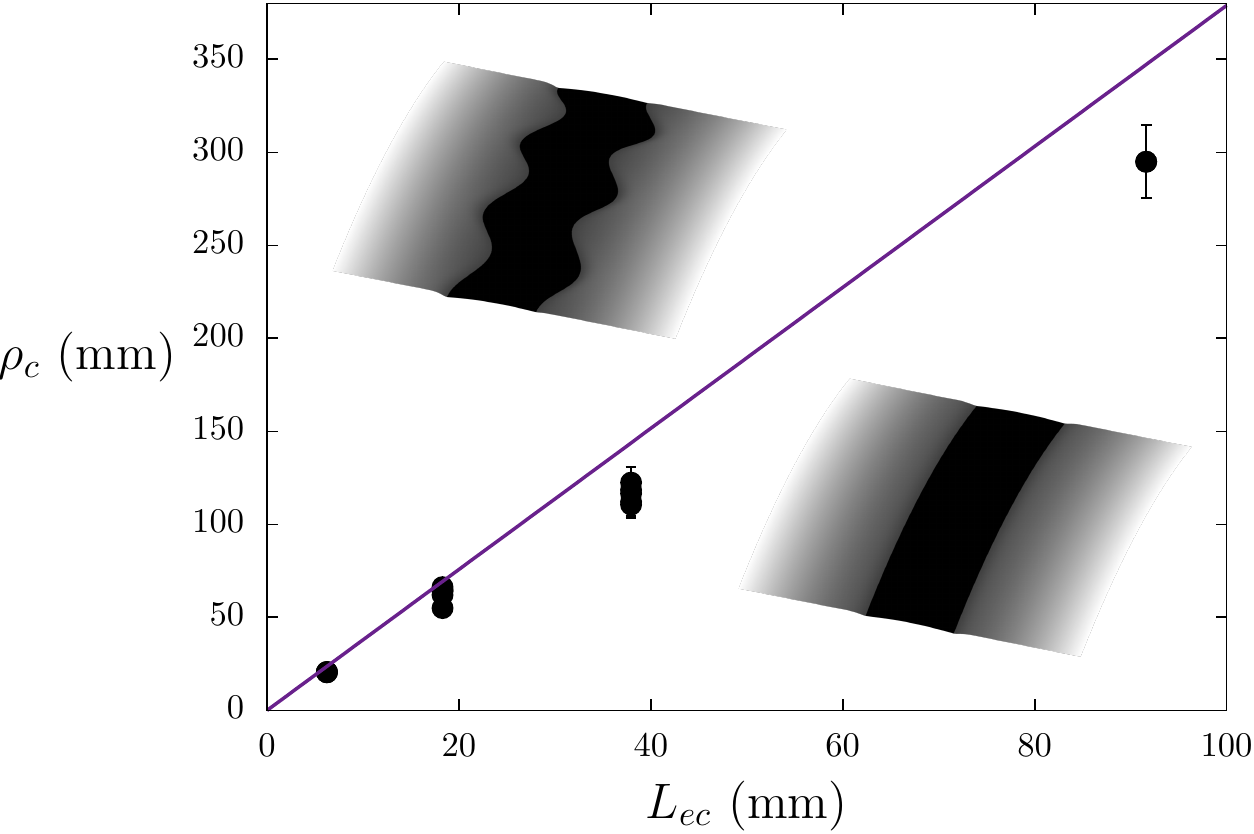}}
    \end{minipage}
    \begin{minipage}[b]{0.49\textwidth}
	\subfigure[]{\includegraphics[height=5cm]{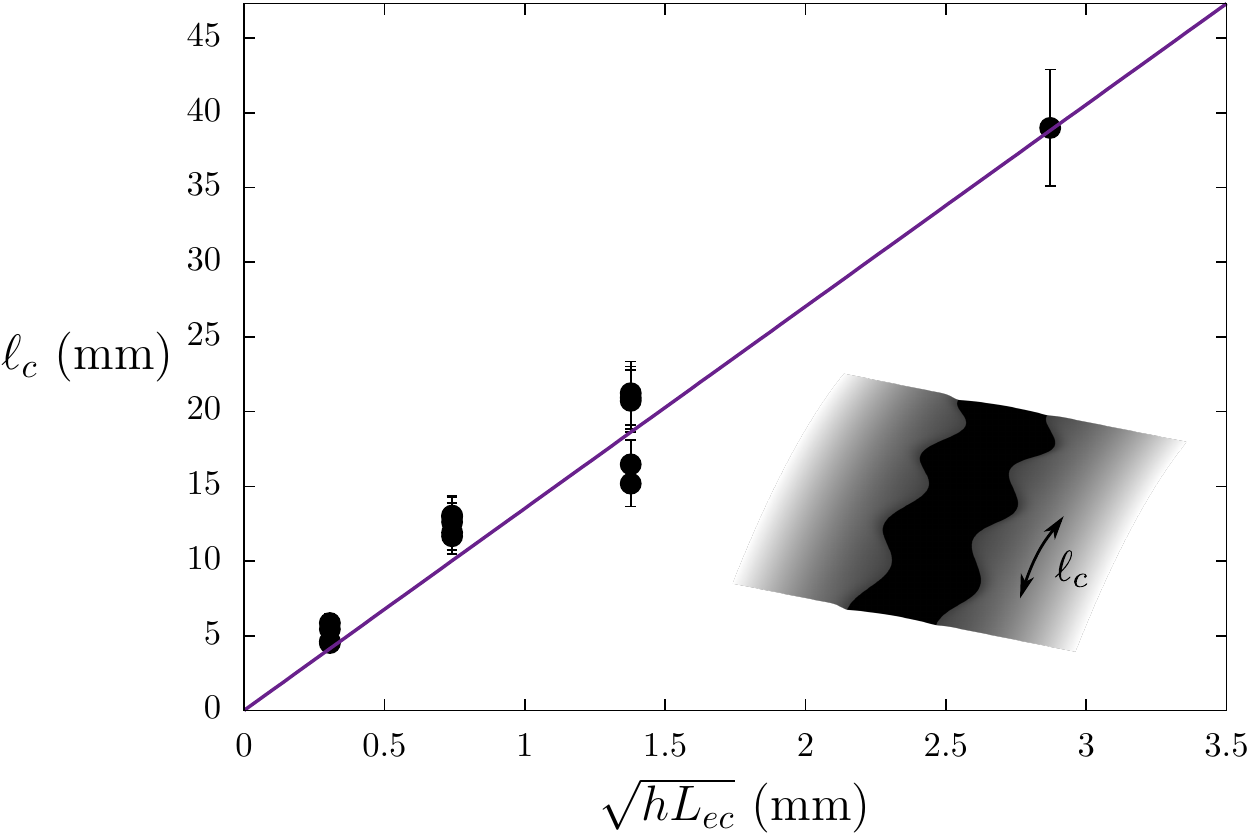}}
    \end{minipage}
    \caption{(a) Critical radius of sphere $\rho_c$ leading to
    undulations of the contact region, as a function of the
    elastocapillary length $L_{ec}$.  Each data point corresponds to a
    specitif film thickness, and is obtained by detecting the
    threshold of instability in the experimental curve for
    $a_{1}(\rho)$, as in figure~\ref{fig:setup}b.  The various data
    points correspond to thicknesses $h$ ranging from $15~\mathrm{\mu
    m}$ to $90~\mathrm{\mu m}$.  Note the collapse of these points
    onto a single curve.  (b) Wavelength of the oscillations $\ell_c =
    2\pi/k_c$ at threshold ($\rho=\rho_c$).  In (a) and (b), the solid
    lines show to the prediction of the analysis of linear stability
    in equation~(\ref{eq:PredictionsForNu=.4exp}) for $\nu=0.4$, with
    no adjustable parameter.}
    \label{fig:res}
\end{figure}

In figure~\ref{fig:res} we compare these predictions to the
experiments.
The elastocapillary length $L_{ec}$ is computed from the experimental
parameters using equation~(\ref{eq:Lec}).  The wavelength of the
pattern agrees very well with the prediction of the analysis of linear
stability; there is no adjustable parameter.  The agreement concerning
the instability threshold is not as good but satisfactory; as apparent
in Figure~\ref{fig:setup}d, the actual threshold is lower by about
$15\%$ than that predicted by theory.  This can probably be attributed
to imperfections in the planarity of the film and in the way that it
is layout onto the substrate, as well as to the finite size of the 
meniscus.

\section{Conclusion}

We have investigated the adhesion of a thin film on a spherical
substrate.  The Donnell theory of nearly cylindrical shells has been
used, taking into account the energy of adhesion with the sphere.  We
have derived the boundary conditions holding at the edge of the moving
boundary between the wet and free regions.  A family of nonlinear
solutions describing the unbuckled configuration of the film has been
constructed.  These solutions, which have cylindrical symmetry, are
indexed by a dimensionless adhesion number $\sigma = \rho/L_{ec}$,
that compares the radius of curvature of the sphere $\rho$ to the
elastocapillary length $L_{ec}$.   As the adhesion number
$\sigma$ increases, a compressive stress builds up along the edge
of the contact zone and makes the film buckle.  We have carried out a
linear stability analysis of the unbuckled solution, taking into
account the motion of the moving boundary between the wet and free regions,
and found of a critical value of the adhesion number $\sigma$ above
which the contact region becomes sinous.  The symmetry of the unstable
mode, as well as the instability threshold and the wavelength found by
the theory are in good agreement with the experiments.

In our system, buckling is driven by the geometric frustration due to
the mismatch between the Gauss curvature of the substrate and that of
the planar film; a related type of buckling driven by geometric
frustration has been investigated in plates undergoing
swelling~\cite{KleinEfratSharoShaping-of-Elastic-Sheets-by-Prescription-of-Non-Euclidean07},
and in annular origami models having a curved
crease~\cite{Dias-Dudte-EtAl-Geometric-Mechanics-of-Curved-2012}.  Our
paper provides a detailed analysis of the initial buckling of our
system.  Far above the buckling threshold, the buckling patterns
evolve into a branched network of bands whose edges are made up by a
series of cusps.  These singularities seem to be connected with the
existence of developable cones and sharp ridges in the free parts of
the film, and are not yet understood in detail.

\newpage

\appendix

\section{A justification of the Donnell equations by formal asymptotic expansion}
\label{app:DonnellJustification}

The Donnell equations for shells are justified by a formal expansion
with respect to a small parameter $\eta$ proportional to the
square-root of the aspect-ratio of the shell.  Assuming that the
transverse displacement is of order $\eta^2$, we derive these
equations from the general non-linear equations for elastic shells
under finite displacement.  The present derivation is mainly given for
pedagogical purposes as a rigorous proof is available
in~\cite{figueiredo1991}.  Note that formal expansions have also been
used to justify plate equations in~\cite{ciarlet}.


\subsection{Transformation}

Let us first define the cylindrical basis vectors
\begin{equation}
    \underline{e}_{1}(\theta) = \left(
    \begin{array}{c}
	0 \\Ê\cos\theta \\ \sin\theta
    \end{array}
    \right)
    ,\qquad
    \underline{e}_{2}(\theta) = \left(
    \begin{array}{c}
        0 \\ -\sin\theta \\ \cos\theta
    \end{array}
    \right)
    \textrm{.}
    \label{eq:referenceCylindricalBasis}
\end{equation}
This basis is such that $\underline{e}_{1}'(\theta) =
\underline{e}_{2}(\theta)$, $\underline{e}_{2}'(\theta) =
-\underline{e}_{1}(\theta)$, and $\underline{e}_{1}(\theta) \times
\underline{e}_{2}(\theta) = \underline{e}_{X}$: the polar axis is the
axis $X$ of the shell in reference configuration.

In the reference configuration, the shell is rolled into a cylinder of
radius $\rho$.  We use Lagrangian coordinates $(x,y)$.  The position in
reference configuration is denoted $\underline{x}$, as shown in figure~\ref{fig:ShellReferenceActual}a,
\begin{equation}
    \underline{x}(x,y) = \rho\,\underline{e}_{1}(\theta)
    + x\,\underline{e}_{x},
    \quad\textrm{where }\theta=\frac{y}{\rho}
    \textrm{.}
    \nonumber
\end{equation}
Note that the metric associated with the set of coordinates
$(x,y)$ is the unit tensor,
$\underline{\underline{\nabla}}\underline{x}^T\cdot
\underline{\underline{\nabla}}\underline{x} =
\underline{\underline{1}}$.

The deformed configuration $\underline{\tilde{x}}$ is defined in terms
of the displacement $(u_{x}(x,y), u_{y}(x,y), w(x,y))$ in cylindrical
coordinates by equation~(\ref{eq:deformedPos}), and is also sketched
in figure~\ref{fig:ShellReferenceActual}b.

The strain in the shell is measured using a membrane strain tensor
$\underline{\underline{e}}$ and a curvature strain tensor
$\underline{\underline{b}}$.  The nonlinear membrane strain
$\underline{\underline{e}}$ is a $2\times 2$ symmetric tensor defined
by the classical formula:
\begin{equation}
    \underline{\underline{F}} = 
    \underline{\underline{\nabla}}\underline{\tilde{x}},\quad
    \underline{\underline{C}} = \underline{\underline{F}}^T\cdot 
    \underline{\underline{F}},\quad
    \underline{\underline{e}} = \frac{1}{2}\,(
    \underline{\underline{C}} - 
    \underline{\underline{1}}
    )
    \textrm{,}
    \label{eq:membraneStrainGeneral}
\end{equation}
where the gradient is taken with respect to the Lagrangian coordinates
$(x,y)$.  The curvature strain tensor is defined by
\begin{subequations}
    \label{eq:curvatureStrainGeneral}
\begin{equation}
    b_{\alpha\beta}
    =
    \underline{\tilde{x}}_{,\alpha\beta}(x,y)\cdot \underline{N}(x,y)
    \label{eq:curvatureTensor}
\end{equation}
where the normal to the shell is defined by
\begin{equation}
    \underline{N}(x,y) = -\underline{\tilde{x}}_{,x}(x,y) \times 
    \underline{\tilde{x}}_{,y}(x,y)
    \textrm{.}
    \label{eq:normal}
\end{equation}
\end{subequations}
The minus sign in the definition of the normal in
equation~(\ref{eq:normal}) makes the normal $\underline{N}$ oriented
in same direction as the radial vector $\underline{e}_{1}(\theta)$ of
the cylindrical basis.  Note that the normal is not a unit normal if
the current configuration is not developable,
$e_{\alpha\beta}\neq 0$.  Since we consider deformations that
are almost inextensible, our definition~(\ref{eq:curvatureTensor}) of
the bending strain is very close to the geometrically exact definition
of the curvature that makes use of a unit normal vector, and this
introduces a higher-order correction in the thin-shell limit.

By inserting equation~(\ref{eq:deformedPos}) into
equation~(\ref{eq:membraneStrainGeneral}), one could rederive the
fully non-linear expression of the 3 independent components
$e_{\alpha\beta}$ in terms of the displacement functions
$(u_{x},u_{y},w)$, relevant to the general theory of shells under
finite displacements.  Similarly, by inserting into
equation~(\ref{eq:curvatureStrainGeneral}), one could derive a fully
non-linear expression for the curvature strains $b_{\alpha\beta}$.

\subsection{Formal expansion of the membrane and curvature strains}

The Donnell equations are derived from the above set of equations
under the assumption of a moderate displacement.  We consider a formal
expansion of the above equations with respect to a small parameter
$\eta$.  The definition of $\eta$ given in equation~(\ref{eq:defEta})
will be justified.  For the moment, it is sufficient to
assume $\eta \ll 1$.  We assume that the transverse displacement $w$
scales like $\rho\,\eta^2$, that the in-plane displacement scales like
$\rho\,\eta^3$, and that the typical scale for the tangent coordinates
$x$ and $y$ is $\rho\,\eta$.  The dimensionless displacement
$\overline{u}_{\alpha}(\overline{x}, \overline{y})$ and
$\overline{w}(\overline{x}, \overline{y})$ are defined in
equation~(\ref{eq:rescaling1}) in terms of the rescaled coordinates
$\overline{x} = x/(\rho\,\eta)$ and $\overline{y} = y/(\rho\,\eta)$.

The previous scalings can be justified as follows.  Our starting
assumption is that the deflection $w$ is small, and scales as
$\rho\,\eta^2$.  From this, as $w \sim x^2/\rho$ along the curved
shell, it appears that the natural scale for the tangent coordinates
is $\rho\,\eta$.  Balancing the linear and non-linear terms in the
membrane strain, we have $u_{\alpha,x} \sim w_{,x}^2$ and so the scale
for the tangential displacement is $\rho\,\eta^3$.

Expanding the membrane strain $\underline{\underline{e}}$ introduced
in equation~(\ref{eq:membraneStrainGeneral}) with respect to $\eta$,
one computes
\begin{equation}
    e_{\alpha\beta}(x,y) = 
    \eta^{2}\,\overline{e}_{\alpha\beta}(\overline{x},\overline{y})
    +\mathcal{O}(\eta^4)
    \textrm{,}\\
    \label{eq:formalExpansionE}
\end{equation}
where the dominant contribution is given by
\begin{equation}
    \overline{e}_{\alpha\beta}(\overline{x},\overline{y})
    =
    \left(
	\frac{
	\overline{u}_{\alpha,\beta}(\overline{x},\overline{y})
	+
	\overline{u}_{\beta,\alpha}(\overline{x},\overline{y})
	}{2}
	+
	\delta_{\alpha y}\,\delta_{\beta y}\,\overline{w}(\overline{x},\overline{y})
	\right)
	+\frac{1}{2}\,\overline{w}_{,\alpha}(\overline{x},\overline{y})
	\,\overline{w}_{,\beta}(\overline{x},\overline{y})
    \textrm{.}
    \label{eq:formalExpansionEDominant}
\end{equation}
This result was stated in equation~(\ref{eq:DonnellMembraneStrain})
but with the bars omitted.  In the right-hand side, the derivatives
are taken with respect to the rescaled variables, for instance
$\overline{u}_{\alpha,\beta} = \partial \overline{u}_{x}/\partial
\overline{y}$ when $\alpha=x$ and $\beta=y$.

A similar expansion of the curvature strain defined in
equation~(\ref{eq:curvatureStrainGeneral}) yields
\begin{equation}
    b_{\alpha\beta}(x,y) = 
    \frac{\eta^0}{\rho}\,\overline{b}_{\alpha\beta}(\overline{x}, 
    \overline{y}) + \mathcal{O}(\eta^2)
    \label{eq:formalExpansionK}
\end{equation}
where the dominant contribution reads
\begin{equation}
    \overline{b}_{\alpha\beta}(\overline{x},\overline{y}) =
    -\overline{w}_{,\alpha\beta}(\overline{x},
    \overline{y}) \textrm{.}
    \label{eq:formalExpansionKDominant}
\end{equation}
The minus sign in the right-hand side comes from that introduced in
the definition of the normal $\underline{N}$.

\subsection{Rescaled constitutive equations, elastic energy}

For an isotropic, Hookean (linearly elastic) material, the
constitutive laws read, in physical units:
\begin{subequations}
    \label{eq:formalExpansionConstitutive}
    \begin{align}
        n_{\alpha\beta} & = C\,(
	(1-\nu)\,e_{\alpha\beta}
	+
	\nu\,(\tr \underline{\underline{e}})\,\delta_{\alpha\beta}
	)
        \label{eq:formalExpansionConstitutive-Membrane}\\
        m_{\alpha\beta} & = D\,
	(
	(1-\nu)\,b_{\alpha\beta}
	+
	\nu\,(\tr \underline{\underline{b}})\,\delta_{\alpha\beta}
	)
        \label{eq:formalExpansionConstitutive-Bending}
    \end{align}
\end{subequations}
where $C=Eh/(1-\nu^2)$ and $D=Eh^3/[12(1-\nu^2)]$ are the stretching
and bending moduli, respectively.  It is convenient to define the
slenderness parameter $\eta$ by
\begin{equation}
    \eta = \left(\frac{D}{C\,\rho^2}\right)^{1/4} = \left(\frac{1}{\sqrt{12}}\,\frac{h}{\rho}\right)^{1/2}
    \textrm{,}
    \label{eq:etaDef}
\end{equation}
as we did earlier in equation~(\ref{eq:defEta}).  Indeed, this
convention makes both the stretching and bending moduli $C$ and $D$
effectively equal to one in rescaled units.

The elastic energy of the shell reads
\begin{equation}
    E_{\mathrm{shell}} = \frac{1}{2}\,\iint(
    n_{\alpha\beta}\,e_{\alpha\beta}
    +
    m_{\alpha\beta}\,b_{\alpha\beta}
    )\,\mathrm{d}x\,\mathrm{d}y
    \label{eq:shellEnergy}
\end{equation}
and can be rescaled as
\begin{equation}
    E_{\mathrm{shell}}  = \rho^2\,C\,\eta^4\,\overline{E}_{\mathrm{shell}}
    \textrm{,}
\end{equation}
where
\begin{equation}
    \overline{E}_{\mathrm{shell}} = \frac{1}{2}\,\iint(
    \overline{n}_{\alpha\beta}\,\overline{e}_{\alpha\beta}
    +
    \overline{m}_{\alpha\beta}\,\overline{b}_{\alpha\beta}
    )\,\mathrm{d}\overline{x}\,\mathrm{d}\overline{y}
    \textrm{.}
    \label{eq:ShellEnergy-Rescaled}
\end{equation}
The dimensionless stress can be computed by the dimensionless version
of the constitutive law, see equation~(\ref{eq:constitutive}).

As shown in section~\ref{eqequi}, Donnell equations then follow by
variational principles from the shell
energy~(\ref{eq:ShellEnergy-Rescaled}) combined with the
definitions~(\ref{eq:formalExpansionEDominant})
and~(\ref{eq:formalExpansionKDominant}) of the membrane and curvature strains,
and with the rescaled constitutive laws~(\ref{eq:constitutive}).
These definitions~(\ref{eq:formalExpansionEDominant})
and~(\ref{eq:formalExpansionKDominant}) are often presented as approximation.
We have just shown that they follow from scaling assumptions and are
asymptotically exact for small displacement.

\section{Two-dimensional Weierstrass-Erdmann corner conditions}
\label{app:CornerConditions}

\begin{figure}
    \centering
    \includegraphics{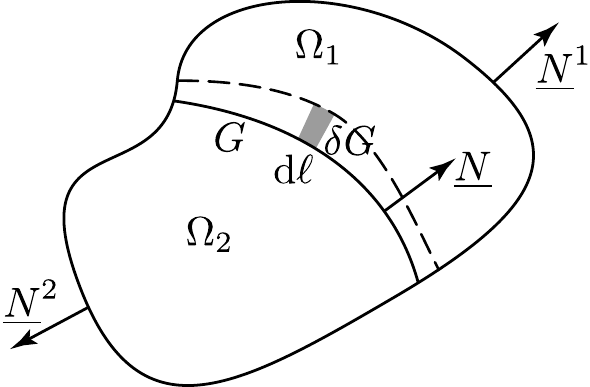}
    \caption{Equilibrium of a mobile interface $G$ between two domains
    $\Omega^1$ and $\Omega^2$.}
    \label{fig:adhesion}
\end{figure}

With the aim to derive the jump conditions at the interface
between the adhering part and the free part of the shell, we recall
the Weierstrass-Erdmann corner conditions in a generic two-dimensional
setting.  We refer to~\cite{troutman} for a detailed presentation.  We
consider a two-dimensional domain $\Omega$ that is split in two
regions $\Omega_1$ and $\Omega_2$ meeting along a boundary curve $G =
\Omega_1 \cap \Omega_2$, as depicted in figure~\ref{fig:adhesion}.
The unknowns are the functions $\xi_{\alpha}(x,y)$ where $\alpha$ is
an index, such as the component of the displacement in the case of an
adhering shell.  Each region $\Omega_{i}$, $i=1,2$ is associated with
a specific Lagrangian (also called energy functional),
\begin{equation}
    E_i(\xi_{\alpha},G) = \int_{\Omega_i(G)}
    \mathcal{L}^i(\xi_{\alpha},\xi_{\alpha,\beta})\,\mathrm{d}x\,\mathrm{d}y
    \textrm{.}
\label{eq:erdmanenergy}
\end{equation}
The boundary may evolve and we use the boundary curve $G$ as an
unknown: the domains $\Omega_i$ are reconstructed in terms of the
curve $G$, hence the notation $\Omega_i(G)$.

We are interested in the conditions that make the total energy $E =
E_{1} + E_{2}$ minimum, and in particular in the conditions associated
with the motion of the interface $G$.  The variation $\delta E$ of the
total energy is made up of a term arising from the variation of the
functions $\xi_{\alpha}(x,y)$ inside each domain, and from a term
associated with the motion of the free boundary,
\begin{equation}
    \delta E = 
    \sum_{i=1}^2
    \int_{\Omega_i(G)}
    \left(
    \frac{\partial \mathcal{L}^i}{\partial \xi_{\alpha}}
    \,\delta\xi_{\alpha}
    +
    \frac{\partial \mathcal{L}^i}{\partial \xi_{\alpha,\beta}}
    \,\delta\xi_{\alpha,\beta}    
    \right)
    +
    \int_{G}
    (\mathcal{L}^2 - \mathcal{L}^1)
    \;\delta G\,\mathrm{d}\ell
    \textrm{.}
    \label{eq:ErdmannStep1}
\end{equation}
Here the partial derivatives denote functional derivative,
$\mathrm{d}\ell$ is the element of length along the boundary $G$, and
$\delta G\,\mathrm{d}\ell$ is the signed area swept by the free
boundary, counted positively when the region $\Omega_{2}$ grows while
the region $\Omega_{1}$ shrinks, as shown in
figure~\ref{fig:adhesion}.

We consider the contribution $\delta E_{G}$ to $\delta E$ that
collects all terms written as integrals along the boundary curve $G$.
These terms yield the jump conditions associated with the
equilibrium of the boundary, while the other terms contribute to the
Euler-Lagrange conditions of equilibrium inside each subdomain.  This
$\delta E_{G}$ is made up of the last term in
equation~(\ref{eq:ErdmannStep1}), and of a term coming from the
integration by part of the term proportional to
$\delta\xi_{\alpha,\beta}$,
\begin{equation}
    \delta E_{G} = \int_{G}\left(
    \discont{\mathcal{L}^i}\,\delta G
    +
    \discont{
    \frac{\partial \mathcal{L}^i}{\partial \xi_{\alpha,\beta}}
    \,\delta\xi_{\alpha}
    }\,N_{\beta}
    \right)\,\mathrm{d}\ell
    \textrm{.}
    \label{eq:variationerdmann}
\end{equation}
Here the double bracket denotes the jump, $\discont{f^i}= f^2 -
f^1$ and $\underline{N} = (N_{x},N_{y})$ is the vector normal to the
boundary $G$, directed towards region 1 as in the figure.  Along the
common boundary $G$, this $\underline{N}$ is equal to the outward normal
$\underline{N}^2$ with respect to the domain $\Omega_{2}$, and is
opposite to the outward normal $\underline{N}^1$ to domain
$\Omega_{1}$.

We shall assume that the unknown $\xi_{\alpha}$ is prescribed to be
continuous across the boundary, as happens with the components of the
displacement and with the slope $q_{\alpha} = w_{,\alpha}$ of an
elastic shell,
\begin{equation}
    \discont{\xi_{\alpha}} = 0
    \textrm{.}
    \label{eq:xiContinuous}
\end{equation}
This continuity relation can be differentiated in a frame moving along
with the boundary.  This yields, for an arbitrary perturbation of the
boundary and of the function,
\begin{equation}
    \discont{\delta \xi_{\alpha}}
    +\discont{\xi_{\alpha,\beta}}\,N_{\beta}\,\delta G = 0
\label{eq:firstordermove}
\end{equation}

First, consider perturbations leaving the boundary unchanged, $\delta
G = 0$.  Then equation~(\ref{eq:firstordermove}) shows that $\delta
\xi_{\alpha}$ is continuous across the boundary as $\discont{\delta
\xi_{\alpha}}=0$.  In the expression for $\delta E_{G}$ given in
equation~(\ref{eq:variationerdmann}), the first term cancels when the
free boundary is at rest, and $\delta \xi_{\alpha}$ can be factored out
of the jump operator, yielding the condition
\begin{equation}
    \discont{
    \frac{\partial \mathcal{L}^i}{\partial \xi_{\alpha,\beta}}
    }
    \,N_{\beta}\,\delta\xi_{\alpha}
     = 0
     \textrm{.}
    \label{eq:eqjj1}
\end{equation}

In the particular case of a function $\xi_{\gamma}$ whose values are
prescribed in the domain $\Omega_{1}$, we have $\delta\xi_{\gamma}^1 =
0$.  The continuity condition~(\ref{eq:eqjj1}) then yields $\delta
\xi_{\gamma}^2 +\discont{\xi_{\gamma,\beta'}}\,N_{\beta'}\,\delta G =
0$.  Cancelling the variation in equation~(\ref{eq:variationerdmann})
then yields
\begin{equation}
    \textrm{if $\delta\xi_{\gamma}$ is prescribed on $\Omega_{1}$, 
    then }\mathcal{L}^2 - \mathcal{L}^1 - 
    \frac{\partial \mathcal{L}^2}{\partial \xi_{\gamma,\beta}}\,
    \discont{\xi_{\gamma,\beta'}}\,N_{\beta'}\,N_{\beta} 
    = 0
    \textrm{.}
    \label{eq:eqjj}
\end{equation}

\bibliographystyle{elsarticle-num.bst}
\bibliography{donnell.bib}

\end{document}